# Report of the ICFA Beam Dynamics Workshop

## *"Accelerators for a Higgs Factory: Linear vs. Circular"*

## (HF2012)

Alain Blondel[1], Alex Chao[2], Weiren Chou[3], Jie Gao[4], Daniel Schulte[5] and Kaoru Yokoya[6]

[1] U. of Geneva, Geneva, Switzerland
[2] SLAC, Menlo Park, California, USA
[3] Fermilab, Batavia, Illinois, USA
[4] IHEP, Beijing, China
[5] CERN, Geneva, Switzerland
[6] KEK, Tsukuba, Japan

# Contents

# 1 Executive Summary

The 4th of July 2012 was a historical moment for high-energy physics (HEP). On that day, CERN announced that both the ATLAS and CMS experiments had discovered a new Higgs-like boson. In this report, we shall assume that this newly found particle will turn out to be a Higgs boson, the key signature of the Standard Model. At the ICFA meeting on July 8 in Melbourne, an ICFA Beam Dynamics Workshop "*Accelerators for a Higgs Factory: Linear vs. Circular*" (HF2012) was approved. This workshop took place from November 14 to 16, 2012 at Fermilab, USA. (conferences.fnal.gov/hf2012) Seventy-one people from 31 institutions in Asia, Europe and North America attended. The workshop agenda is in Appendix 1.

The next "big" collider after the LHC has been the subject of international HEP community planning for more than a decade. For example, at the 2001 Snowmass meeting, a number of options were considered: a linear $e^+e^-$ collider, a circular $e^+e^-$ collider, a Very Large Hadron Collider (VLHC) and a muon collider. Since then, however, under the direction of ICFA, a series of decisions have been made:

- It should be an $e^+e^-$ collider;
- It should be a linear $e^+e^-$ collider;
- It should be a cold (i.e., Superconducting RF) linear $e^+e^-$ collider.

These major decisions have been followed by a number of significant steps. An international collaboration for the ILC, defined as a Superconducting RF linear $e^+e^-$ collider of a center-of-mass energy ($E_{CM}$) 500 GeV was formed under the leadership of the Global Design Effort (GDE). Substantial progresses in design, prototyping and R&D have been made. A Technical Design Report (TDR) will be published in mid-2013, complete with a cost estimate. The Japanese HEP community has proposed to build a low-energy linear collider ($E_{CM}$ = 250 GeV) as the first step of the ILC.

The discovery of the Higgs-like boson at the LHC has placed the focus on the need to study the properties of this new particle with high precision. Thanks to nature, the light mass of the Higgs (~126 GeV) puts a Higgs factory closer to reach. Various proposals are claimed to be able to reach the energy and luminosity of interest for a Higgs factory: a linear $e^+e^-$ collider, either cold or warm (e.g., x-band), a circular $e^+e^-$ collider; a muon collider and a photon collider. They were put on the table at the HF2012 workshop as possible candidates:

(a) Linear $e^+e^-$ colliders:
- ILC
- CLIC
- X-band klystron-based

(b) Circular $e^+e^-$ colliders:
- Fermilab site-filler
- LEP3
- TLEP
- SuperTRISTAN-40 and SuperTRISTAN-80
- CHF-1 and CHF-2
- VLLC

(c) Muon collider

(d) Photon colliders:
  - ILC-based
  - CLIC-based
  - Recirculating linac-based (SAPPHiRE)
  - SLC-type

The purpose of the workshop was to compare the pros and cons of these candidates mainly from the accelerator point of view but not to recommend any specific machine, which is only possible with further input from the physics side. Also it is not excluded that multiple facilities may be required to complement one another in addressing the entire breadth of Higgs physics. The comparison includes:

- physics reach
- performance (energy and luminosity)
- upgrade potential
- technology maturity and readiness
- technical challenges requiring further R&D

Cost was not included because it was too early for such a comparison at this stage. Parameter comparison tables are in Appendix 2. These tables were provided by the workshop presenters except for some obvious corrections and items derived from the data provided.

The LHC will keep collecting valuable data and it is expected that more data from the LHC will further clarify what kind of Higgs factory (or factories) will be needed.

For the candidates above (except the muon collider operating at the Higgs resonance) delivered luminosities are in units of $10^{34}$ $cm^{-2}s^{-1}$. Assuming $1\times10^{7}$ seconds a year for effective machine operation, this leads to an integrated luminosity of 100 $fb^{-1}$ per year. Since the cross-section for Higgs production (the ZH channel for $e^+e^-$ and s-channel for $\gamma\gamma$) is about 200 fb, the "quantum" is ~20,000 Higgs per year. The muon collider operating at the resonance has a lower luminosity ($10^{31}$-$10^{32}$ $cm^{-2}s^{-1}$). But the cross-section of $\mu^+\mu^- \rightarrow H$ is much larger (~41 pb), which could compensate for the lower luminosity and give a comparable Higgs production rate.

Before we compare linear and circular $e^+e^-$ Higgs factories, it will be useful to review two $e^+e^-$ Z-factories operating at $E_{CM}$ around 91 GeV – the circular LEP ($2.4\times10^{31}$ $cm^{-2}s^{-1}$), and the linear SLC ($3\times10^{30}$ $cm^{-2}s^{-1}$). Both were successfully designed, constructed and operated, and both achieved important physics results: 0.5 million Z decays at the SLC over 11 years and 4 million in each of four experiments at the LEP over 7 years. The exquisite energy calibration at the LEP by resonant depolarization led to 2 MeV precision on the Z mass and width. The single most precise determination of the electroweak mixing angle $sin^2\theta_{eff}^{lept}$ at the SLC was due to 80% longitudinal electron polarization.

The SLC was the first (and only) linear collider ever built. Since then, extensive studies on linear colliders have been carried out with an impressive level of detail. The ILC and CLIC are two flagship programs. The former is based on Superconducting RF technology while the latter on two-beam acceleration. Hundreds of millions of US dollars (or equivalent) have been invested in these programs. Linear colliders are extremely challenging and complex machines, but the key technologies are claimed to be in hand. There exist well-organized international collaborations. The linear collider community will soon be reorganized by combining the ILC, CLIC and detectors together in a single organization – the Linear Collider Collaboration.

In terms of readiness, the linear $e^+e^-$ Higgs factory is clearly the front runner. The main difficulty comes from the high cost of the project. Recently the Japan HEP community issued a report advocating building a 250 GeV ($E_{CM}$) linear collider in Japan as the first stage of the ILC serving as a Higgs factory.

An x-band klystron-based linear collider was previously studied in great detail in both the US and Japan. But the work was stopped after the ICFA selected SRF technology for the ILC in 2004. However, the interest in an x-band linear collider was renewed at CERN, KEK and SLAC during the discussion of a Higgs factory because its cost would be lower than a CLIC at $E_{CM}$ = 250 GeV, and one could add a CLIC section later for an extension to higher energy.

In addition to these linear e+e- colliders, another approach not discussed at the workshop is the use of plasma (beam- or laser-driven) or dielectric wakefields to provide the acceleration field. These technologies are not mature today and face important technical challenges that require vigorous R&D, but have a large potential for Higgs factories and beyond. A plasma scheme has been published at PAC 2009 (p. 2688). An update will be presented at IPAC 2013. A comparison of this technology with other technologies for linear collider Higgs factory is a subject to be considered in the future.

Contrary to the three different technologies (SRF, two-beam and x-band) for a linear $e^+e^-$ collider, all circular $e^+e^-$ collider are similar. The only difference is their size. From the Fermilab site-filler (16 km) to LEP3 (27 km) to TLEP/SuperTRISTAN/CHF (40-80 km) to the VLLC (233 km), they share a number of common features.

The main limitation of a circular $e^+e^-$ collider is that its energy is limited by synchrotron radiation ($P_{SR} \propto E^4$) and thus has no potential for an energy upgrade. (The linear colliders, on the other hand, have advocated an energy up to 3 TeV.) However, a circular $e^+e^-$ collider could be converted to a *pp* collider in the future as the next energy frontier. This was discussed for instance in a plan presented by IHEP, China. It was proposed to begin by constructing a circular 50-70 km circumference tunnel. In Phase 1, the tunnel would host the China Higgs Factory (CHF), a 240 GeV $e^+e^-$ collider. Then in Phase II (20-30 years in the future), when the high field superconducting magnet technology has further matured, a 100 TeV *pp* collider could be built in the same tunnel. A noticeable feature of this plan is that it completely bypasses the ILC option and focuses the physics on either low energy (Z, W and Higgs boson studies) or very high energy (new frontier).

A main advantage of a circular $e^+e^-$ collider of sufficiently large size is to offer a higher luminosity than a linear one at 240 GeV and below. Also, a circular collider can accommodate more than one interaction point (e.g., LEP had 4 IPs). Dozens of circular $e^+e^-$ colliders have been built and operated in the past five decades. The technology is mature and the experience rich.

However, to meet the required energy (240 GeV) and luminosity (a few times $10^{34}$ $cm^{-2}s^{-1}$) of a Higgs factory, several new major technical challenges need to be met. Following the model of the B-factories and synchrotron light sources, all circular $e^+e^-$ colliders adopted an additional circular accelerator as a full energy continuous injector (top-up injection). Due to high beam intensity and small beam size, beamstrahlung (synchrotron radiation of individual particles in the opposing beam's field) will further limit the beam lifetime. Managing this effect requires both the RF system and the machine optics (in both the arcs and the interaction region) to have large momentum acceptance (ranging from 2% to 6% depending on specific proposals). This is a nontrivial design challenge, especially in view of the equally important requirement on the optics for small emittance.

High synchrotron radiation power is another major challenge. The energy loss per turn in these machines is on the order of 10 GeV, and the beam current on the order of 10 mA. These translate to ~100 MW radiation power and must be replenished by the RF system. The major concerns are the RF power coupler, radiation shielding, radioactivation and the required wall power. Even with a 50% wall plug efficiency (which is several times higher than today's most efficient accelerator, the PSI cyclotron), one would need ~200 MW to compensate for synchrotron radiation. This is higher than the current power consumption at CERN (183 MW) and almost four times as large as that at Fermilab (58 MW). The total site power, including cryogenics, magnets, water cooling and injectors would be even higher. At a given energy, this and many other problems are easier in the machines of larger circumference.

Compared to a circular $e^+e^-$ collider, the muon collider has the advantage of no synchrotron radiation or beamstrahlung problem due to the muon mass 207 times more than an electron. This means a muon Higgs factory would be much smaller (a circumference of 0.3 km). Even a TeV-scale collider could be accommodated in an existing campus (e.g., Fermilab). Moreover, the cross section of the s-channel resonance $\mu^+\mu^- \rightarrow H$ is about 40,000 times larger than that of $e^+e^-$ ($\sigma_0$ = 41 pb), which provides a unique way for detailed measurement of the Higgs line shape should it be an unconventional resonance. This requires a relative beam energy spread commensurate to the expected Higgs boson width of 4 MeV – a few times $10^{-5}$! The technological challenges are enormous, in particular the required 4D and 6D ionization cooling of muon beams. Simulations of the cooling processes have made substantial progress, but experimental verifications are lacking. Some critical issues (e.g., RF breakdown in a strong magnetic field) are being addressed and progress has been made towards viable solutions. The MICE international collaboration in the UK and the MAP program in the US are tackling these issues.

Photon colliders were first suggested as possible extensions of two proposed linear colliders (SLC and VLEPP). Photon colliders are based on Inverse Compton Scattering (ICS) by shooting a low energy (~1 eV) laser beam into a high energy (10s of GeV) electron beam to generate a back-scattered high energy (10s of GeV) photon beam for collisions. The advantage is that the cross section for $\gamma\gamma \rightarrow H$ is large and comparable to $e^+e^- \rightarrow ZH$ (~200 fb) but the required energy is much lower (63 GeV for a photon beam, corresponding to 80 GeV for an electron beam, compared to 120 GeV per electron beam in an $e^+e^-$ collider). This makes a photon collider an attractive option for either a low energy linear collider (80 GeV per electron beam) or a low energy circular collider (80 GeV per beam). Furthermore, for a photon collider there is no need for positrons and only one damping ring is needed. However, the physics of a photon collider is not as comprehensive as a 240 GeV $e^+e^-$ collider. There are also machine design issues (e.g., IR optics and removal of the spent electrons) that need to be addressed.

Several photon collider proposals were presented at the workshop. One example is ILC-based, another example CLIC-based. Their bunch structures are very different. The former has long bunch trains (727 μs) and large bunch spacing (554 ns), whereas the latter has short bunch trains (177 ns) and small bunch spacing (0.5 ns). This leads to drastically different requirements for the laser. With large spacing an optical cavity can be used, whereas a short train makes the use of a single laser shot possible. Thanks to the newly formed ICFA-ICUIL collaboration, the study of the required lasers received strong participation from the laser community. For example, the Lawrence Livermore Lab is working on a fusion project LIFE, which will use 384 laser beams for fusion ignition. Just one of these 384 lasers would be sufficient for a warm linac-based (CLIC or x-band) photon collider. However, further R&D is required for either bunch spacing scenario.

Based on the LHeC study at CERN, it was proposed to use an 80 GeV recirculating linac for a photon collider (SAPPHiRE). SLAC, on the other hand, proposed an SLC-type photon collider that uses a single linac (room temperature or superconducting) to accelerate both beams to 80 GeV.

One issue considered to be essential for inclusion in this workshop report is to give a timeline for these proposals. This is a particularly delicate exercise that the HF2012 program committee is undertaking with great hesitation. Not all proposals are at the same level of technical readiness and this is reflected in the uncertainty in the timeline. Although the workshop has addressed in depth the technical readiness of various proposals, many more issues in addition, e.g. cost, available manpower, funding profile, and international setting, will all have great impact on the timeline. Using the best information available the committee generated the timeline of various proposed Higgs factories. These are in Appendix 3.

The study on Higgs factories will continue. The ILC has finished the TDR. The CLIC has completed the CDR and is proceeding to the TDR stage. But design reports for other proposals (circular $e^+e^-$ colliders, muon collider and photon colliders) have not yet been produced. The proponents are encouraged to move from the parameter design to conceptual design with the goal of publishing a report.

This workshop provided a useful and convenient platform for the international community to meet and discuss issues of mutual interest related to a future Higgs factory. Thus, it will be continued. The next workshop is expected to take place in late 2013. The exact dates and venue are yet to be decided.

# 2 Higgs Physics

Is the new particle discovered at the LHC the Higgs boson of the Standard Model? Or does it carry evidence for physics beyond? High precision measurements of the properties of this new particle are the next step. The session on Higgs physics [1] began with a presentation of the LHC achievements in the discovery of the Higgs boson candidate [2-3] and the present status of measurements of its properties. This was followed by an estimate of what measurements can one reasonably expect to come from LHC after the “nominal” run of 300 $fb^{-1}$ at a center-of-mass energy of 14 TeV, or, after the high luminosity upgrade, to 3000 $fb^{-1}$ (HL-LHC) [4-5]. The theoretical introduction described the role of the Higgs boson as a clear instrument and signal of symmetry breaking of the Standard Model and its minimal but somewhat “ad-hoc” nature, as well as the variety of scenarios that have been advocated to introduce it perhaps more naturally. This calls for a variety of measurements of Higgs properties and of other electroweak quantities. A number of scenarios for New Physics Beyond the Standard

Model have been given in the literature [6-8], but a clear ansatz of the precision required to achieve their detection is still work in progress. The “Higgs factories” are accelerators-detector facilities that can do precisely that, and the following were addressed:

- the LHC itself, both “nominal mode” and HL-LHC
- the $e^+e^-$ colliders, either linear collider projects (ILC, CLIC) or the more recent proposals of circular machines operating in the range $m_Z < E_{CM} \leq 350$ GeV
- the $\mu^+ \mu^-$ collider operating at $E_{CM} = m_H$
- a $\gamma\gamma$ collider operating just above $E_{CM} = m_H$.

## 2.1 Physics Case

The Higgs boson candidate is such a special particle that it should be studied in all possible ways. However, there is still much work to do to understand quantitatively the physics case for building a Higgs Factory given that the theorists are only starting to answer the question “*how precisely should Higgs properties be measured*?” With available information effects of New Physics at the TeV scale could be of the order of a few percent, maybe up to O(5%) on Higgs couplings. A discovery (i.e. 5$\sigma$ observation of a discrepancy with the Standard Model prediction) would then require precision well below a percent. How to compare or combine the information given by the LHC, $e^+e^-$ colliders at ZH threshold, $e^+e^-$ colliders at high energy and a $\mu^+\mu^-$ or $\gamma\gamma$ collider is still in its infancy.

A caveat to the physics discussion is in order: The main purpose of the workshop was to initiate an in-depth accelerator discussion. Nevertheless a number of important new results on the precision achievable for Higgs couplings were presented. At the same time, significant differences of views were highlighted – the most important ones are described in the following sections. It is clear that the physics discussion was only a beginning, and will need to be continued in a dedicated, broader framework with participation of experts from all proposals.

## 2.2 The LHC as a Higgs Factory

It was certainly one of the highlights of the meeting that projections showed potential of HL-LHC to reach percent precision. The cross-sections for Higgs production, shown in Figure 2.1 are very large (20 pb at 8 TeV, and increase by substantial factors when going up in energy). Numbers would be even better for the high energy option HE-LHC in the LHC tunnel with new magnets allowing 33 TeV $E_{CM}$, or for the super-high energy option SHE-LHC in a new, 80 km tunnel, allowing to reach 100 TeV, especially for the determination of the Higgs self-couplings as the double Higgs production increases by a factor 9 when going from 14 to 33 TeV. Assumptions on available luminosity and scaling of systematic errors were discussed extensively. LHC experiments so far have actually performed significantly better than expected.

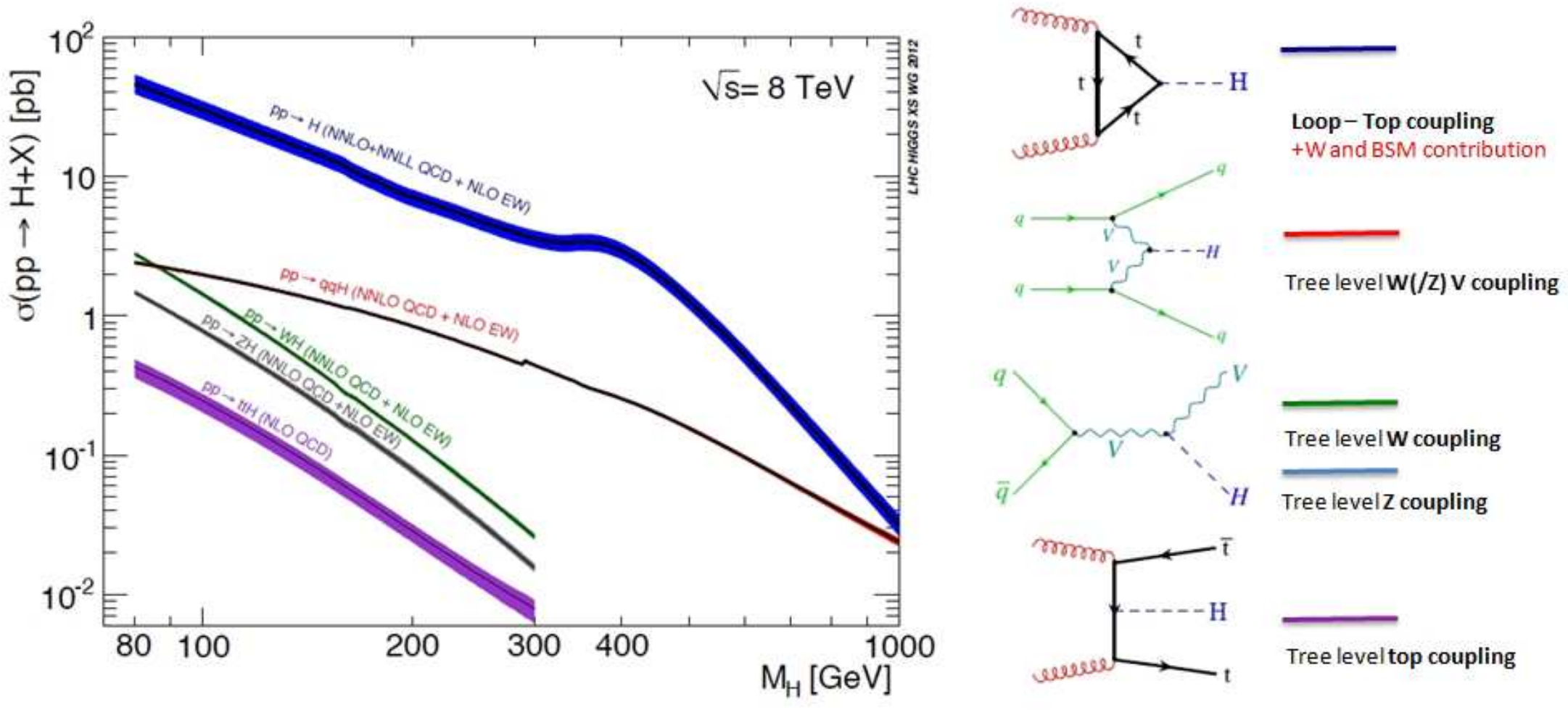

**Figure 2.1**: Higgs production cross-section at the LHC operating at 8 TeV center-of-mass energy. On the right are indicated the five main production mechanisms, which upon tagging can provide coupling measurements to the gluon, W, Z and top.

LHC measurements address the mass, the spin-CP properties and couplings to bosons {gamma, gluon, W, Z} and fermions (top, b, tau, muon) to a few percent. The accessible channels represent $B_{vis}$≈98% of the SM Higgs decays. The spin-CP question is expected to be settled with data available by the end of 2012. The LHC observables are cross-sections times branching ratios and can be expressed in terms of new physics as such

$$O_{if} = (\sigma \times BR)\,\{pp \to ii \to H \to ff\} = \varphi(pp \to ii)\,\frac{\sigma_{ii\to H}\Gamma_{H\to ff}}{\Gamma_H} \equiv O_{if}{}^{(SM)}\ \frac{\kappa_i \cdot \kappa_f}{\kappa_H} \quad ,$$

where: $\kappa_I$ , $\kappa_f$ , $\kappa_H$ represent the ratio between the true value and the Standard Model value for initial and final state coupling to the Higgs and for the total Higgs width; and $\varphi(pp \to ii)$ is the flux factor representing the probability of finding the initial state within a pp collision of the given energy. For a selected final state, the initial state can be identified by kinematic selection. The systematic errors originate from i) various detector efficiencies and performance which tend to improve with increasing statistics, and ii) uncertainties in the flux factors.

In addition, the LHC experiments have come to the preliminary conclusion that, by identification of final states with two Higgs boson decays, an indicative measurement of the effect of the Higgs self-coupling could be obtained with HL-LHC with a precision of the order of 30% or better.

The LHC measurements provide many $O_{if}$ observables with i= gluon-fusion, Vector-Boson-Fusion, or radiation from top, W or Z, for 7 different final states. Phenomenological predictions should be compared directly to these observables. For the sake of comparison with other facilities, projections were made in terms of κ factors, see Table 2.1, with $\Delta\kappa_x \equiv \frac{\Delta g_{Hxx}}{g_{Hxx}}$ to match the $e^+e^-$ notations. In effect what LHC is able to determine very precisely are ratios of branching fractions. What cannot be determined is the common factor $\kappa_H$ that would affect the total width, and could be changed in a global way by new undetected Higgs boson decays. For instance an increase of the total width stemming from a global factor increasing all couplings would result in an increase of the observed cross-sections, while a similar increase due to a significant invisible width would decrease them – conceptually a conspiracy could take place. The power of the variety of initial states available at the LHC is illustrated by the precisions with which the ttH coupling or gluon-gluon Higgs couplings can be determined at the LHC.

## 2.3 Higgs Physics of Electron-Positron Colliders

The most studied Higgs Factory is the $e^+e^-$ collider. Most studies have been made of linear colliders [9-12], but the Higgs physics in $e^+e^-$ depends only marginally on the fact that the collider is linear or circular [13-14]. It then boils down to the availability of high luminosity at the desired energies.

The specific feature is presented by the $e^+e^- \to Z\,H$ reaction, shown in Figure 2.2. The events can be detected inclusively, independently of the Higgs decay mode, by tagging e.g. a leptonic Z decay with a recoil mass equal to the Higgs mass. This cross-section is proportional to $g^2_{HZZ}$, while the cross-section for $e^+e^- \to Z\,H, H \to ZZ$ is proportional to $g^4_{HZZ}/\Gamma_H$, thus allowing the Higgs total width to be determined (this assumes a single resonance). In addition, the investigation of tagged ZH events can reveal invisible or exotic decays that would have escaped detection at the LHC, thus removing the ambiguity between new physics in couplings or in new decay modes. The best place to study this reaction is just below the cross-section maximum of 200 fb at $E_{CM}$ ~ 240(±10) GeV; the chosen value depends on the energy dependence of the luminosity in a given machine. Thus in circular machines the lower end of the bracket is preferred, in linear colliders the upper end. This reaction can also be investigated at any energy where enough statistics can be collected, so that the CLIC studies [11-12] concentrate on the higher energies from 350 GeV up.

For the study of HZ, beam polarization is not essential; a high level of longitudinal electron polarization combined with 30% positron polarization can be arranged to provide a 30% increase of the rate; this was taken into account in the ILC estimates. At 240 GeV the unpolarized cross-section is 200 fb, so that collection of a million ZH events requires an integrated luminosity of 5 $ab^{-1}$. In most $e^+e^-$ Higgs studies the precision is limited by the available statistics. At the linear collider, the studies of the $b\bar{b}$ and $c\bar{c}$ decays benefit from the very small beam spot size at the IP and from the specific time structure of the accelerator that allows operation of the vertex detectors in a pulsed mode. Whether this quality can be preserved on a circular machine with a collision rate between 10 kHz and 1 MHz and in the presence of synchrotron radiation needs further study.

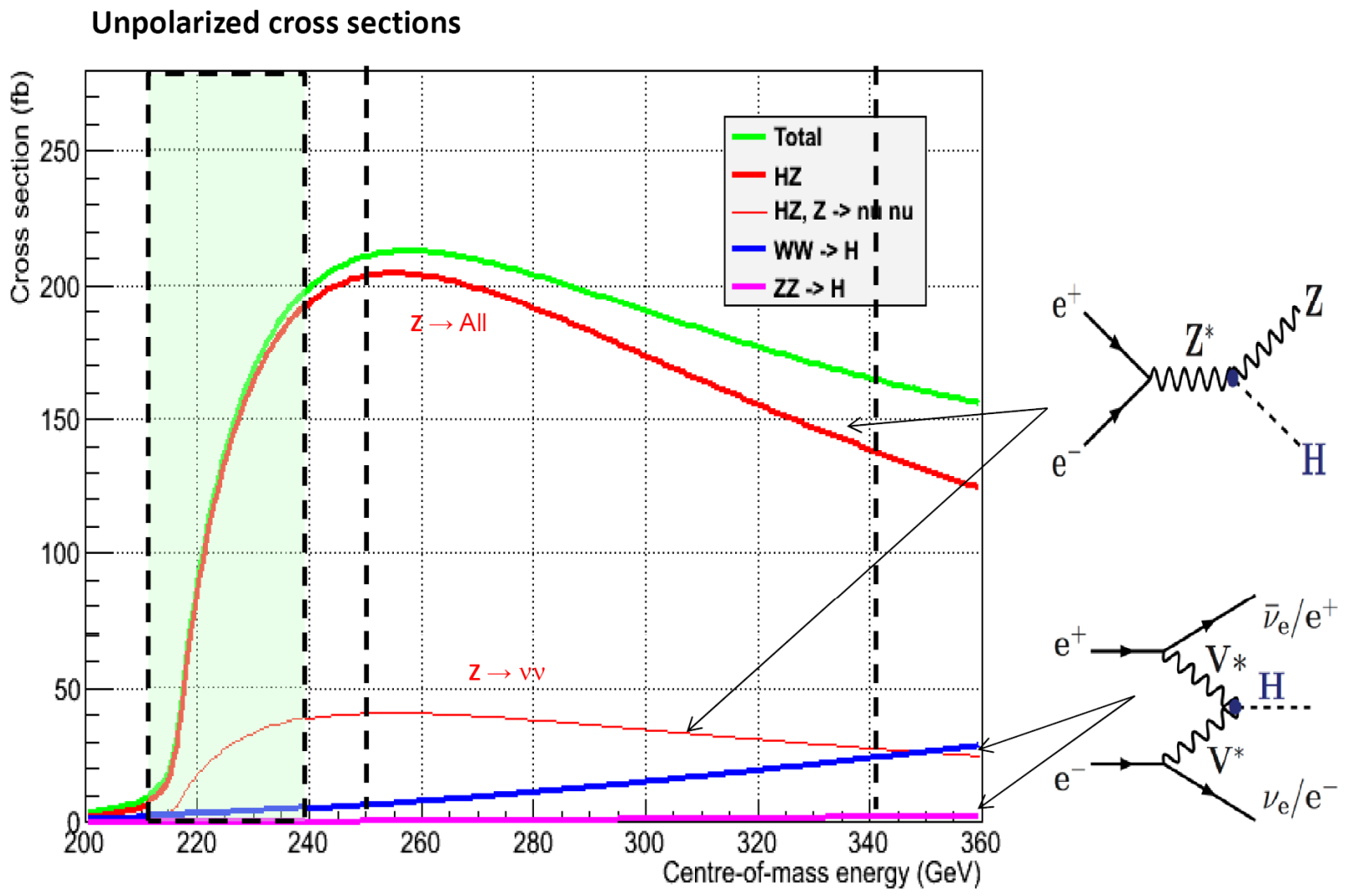

**Figure 2.2:** Unpolarized Higgs boson production cross-sections at low energy $e^+e^-$ colliders. The regions of interest for the direct study of the Higgs boson are: 1) the threshold region, between 210 GeV and 240 GeV; 2) the region of cross-section maximum around 240-250 GeV and the regieon up to the $t\bar{t}$ threshold (340-350 GeV).

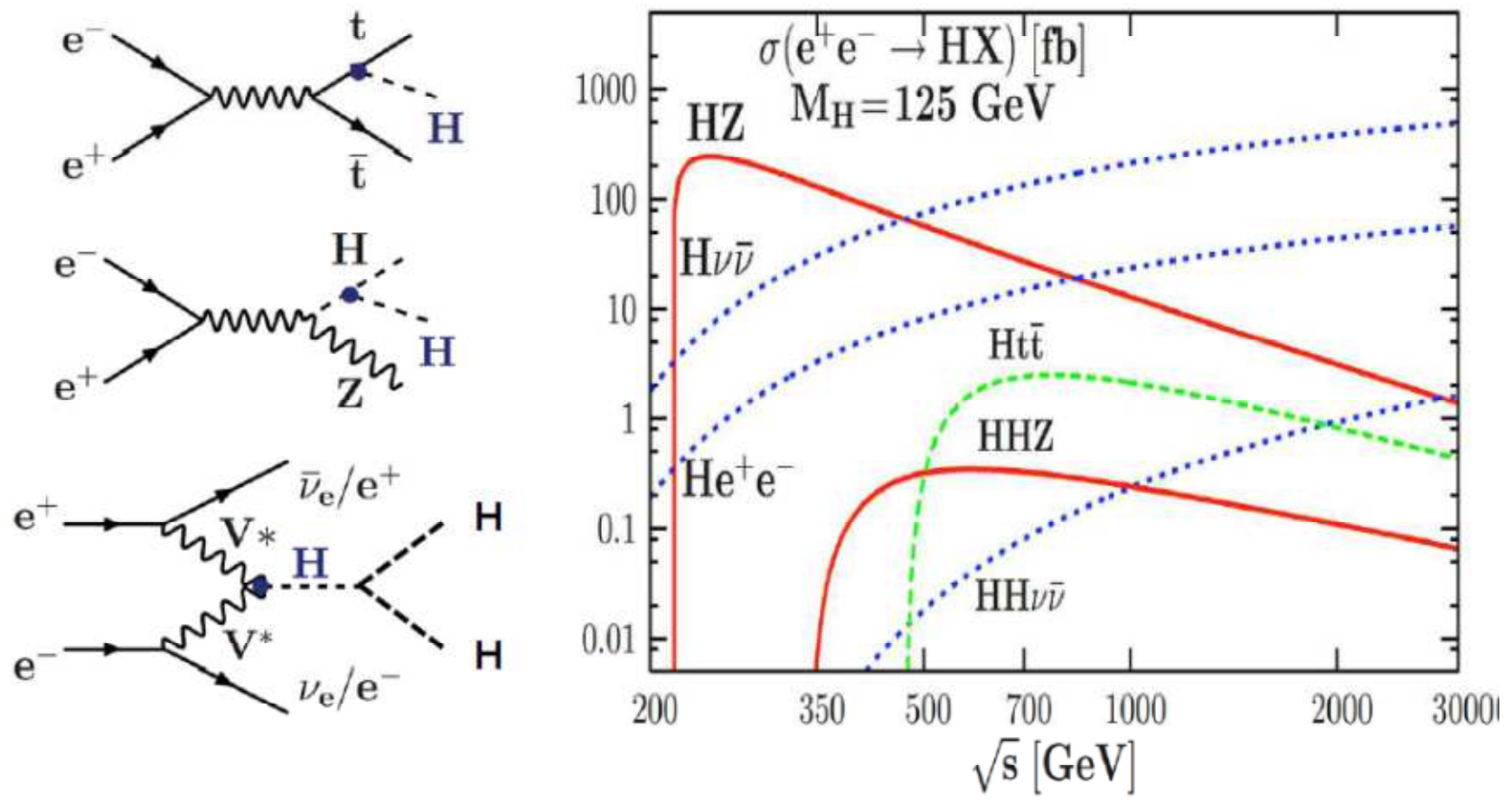

**Figure 2.3**: High energy $t\bar{t}$H and HH cross-sections in $e^+e^-$ colliders. Note the two to three orders of magnitude lower cross-sections compared to the ZH process.

The $e^+e^-$ colliders can also collect high statistics of Higgs decays at higher center-of-mass energies, where the most abundant production mechanism is Vector-Boson Fusion $e^+e^- \to \overline{\nu_e} W^+ \nu_e W^-$, $W^+W^- \to H$ , Figure 2.3. This reaction in combination with the ZH measurements can improve significantly the determination of the Higgs width. At high energies, the reactions $e^+e^- \to t\bar{t}H$ (above 475 GeV) and $e^+e^- \to \overline{\nu_e}\nu_e HH$ (in the TeV range) can give a handle on the Higgs coupling to the top quark and on the Higgs self-coupling. Here the longitudinal beam polarization is useful (although not essential) as a means of controlling the backgrounds and enhancing the signal.

There are different views on the need for an $e^+e^-$ Higgs factory to run at energies higher than the ZH maximum. Everyone agrees that reaching 350 GeV $E_{CM}$ is a definite bonus, allowing the study of the reaction $e^+e^- \to \overline{\nu_e}\nu_e H$ and of direct top quark pair production. This is possible for the ILC, CLIC and, although close to their limit, for the larger circular machines such as CHF and TLEP. The relative merits of running above 350 GeV *for Higgs physics alone* depend on the machine considered. While the linear collider proponents emphasize that the study of the Higgs-top coupling and the triple Higgs coupling require an $e^+e^-$ machine at least up to 1 TeV, the proponents of circular machines argue that these measurements will in all likelihood be performed, earlier, at the HL-LHC with a precision similar to what is advertised for, e.g., the 1 TeV ILC – this was rather new information at the time of the workshop. Running only above 350 GeV as advocated by CLIC proponents is another possible strategy that remains to be fully evaluated. The Higgs self-coupling is an extremely difficult measurement, the best reported sensitivity (very preliminary, 11%) would be from CLIC running at 3 TeV for 2 $ab^{-1}$. A further physics case beyond H(126) for a Linear Collider at an energy above 350 GeV may come from the discovery of one or more new particles at the LHC in the coming years.

In addition $e^+e^-$ colliders are unique for the precision measurements of quantities sensitive to new particles through electroweak radiative corrections (EWRCs). These provide important tests of the completeness of the Standard Model and of the Higgs mechanism. This can be best done by revisiting the Z peak with a high luminosity machine. TLEP claims $10^{36}$ $cm^{-2}s^{-1}$ (TeraZ). The main question there is how one could take advantage of these potentially huge statistics to improve on the LEP measurements, some of which are already at the limit of systematics.

In this context, the availability of longitudinal beam polarization is extremely valuable for the measurement of the inclusive and exclusive beam polarization asymmetries $A_{LR}$ and $A^f_{FB,Pol}$. This is straightforward in a linear collider and has been advocated for obtaining a measurement of the weak mixing angle $\sin^2{}_{eff}{}^{lept}$ with a precision of a few $10^{-5}$. If longitudinal polarization could be achieved with colliding beams in a circular machine at the Z peak, a small fraction of the advertised luminosity would allow these measurements to reach an extremely interesting level of accuracy.

The availability of very precise energy calibration, as available from the transverse polarization in a circular machine using resonant depolarization, would allow a new measurement of the Z mass and width with ten-fold precision improvement, or better, over the present errors of about 2 MeV. The W pair threshold would offer a measurement of the W mass with better than 1 MeV precision if a measureable level of transverse polarization can be achieved at ~80 GeV per beam.

## 2.4 Physics of $\mu^+\mu^- \rightarrow$ Higgs

A muon collider [15-18] can do everything that an $e^+e^-$ collider of the same energy can do, with some advantages in terms of precise knowledge of the center-of-mass energy distribution, which can be extracted exquisitely from (g-2) spin precession detected from the decay electrons. However, a $\mu^+\mu^-$ collider has the additional feature that the coupling of muons to the Higgs is $m_\mu/m_e$ times larger than for electrons, leading to a useable cross-section of 40 pb for the s-channel production $\mu^+\mu^- \rightarrow H(126)$. The study of the resonance requires a machine of precisely $E_{beam} = m_H/2$ with a precision of better than $\Gamma_H = 4.2$ MeV. The demand on both reproducibility and beam energy spread is thus very stringent. The energy spread can in principle be reduced to $3\times10^{-5}$ by emittance exchange but this is done at the expense of transverse emittance and the luminosity is expected to be around $10^{31}$ $cm^{-2}s^{-1}$.

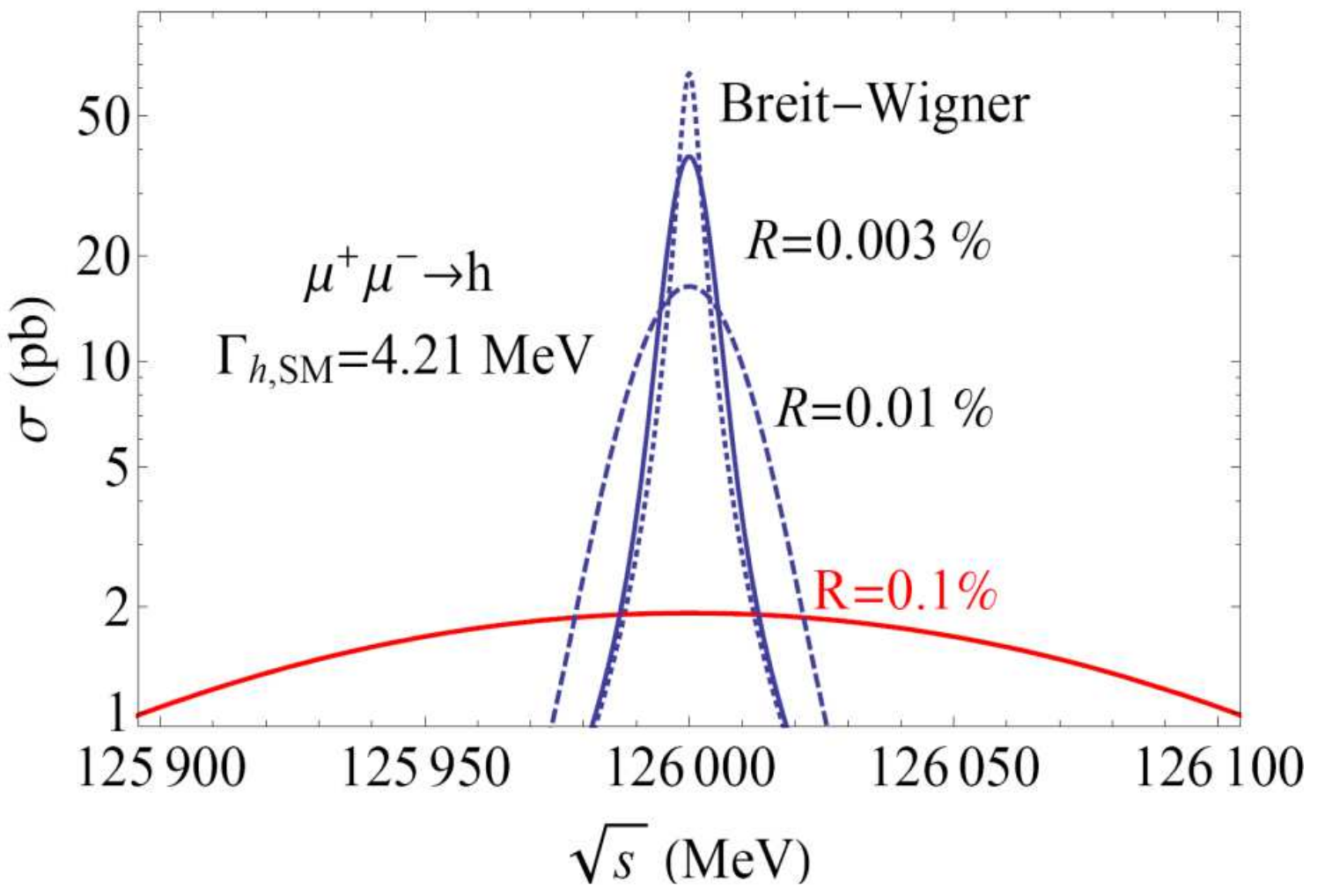

**Figure 2.4:** Direct measurement of the Higgs decay width Γ by using muon beams of high energy resolution.

About 2300 Higgs decays would be observed for one year of running (100 $pb^{-1}$). Then the Higgs boson mass can be obtained with a precision of 0.1 MeV, the line shape width directly with a precision of 0.2 MeV (5%), the peak cross-section with a precision of 2-3%, from which the Higgs width can be extracted with a relative precision of 3% and the muon coupling to 1.5%.

In conclusion, the muon collider Higgs factory is unique in its ability to check that the Higgs boson is a single resonance of the expected line shape; the Higgs mass can be determined with a precision of a 0.1 MeV, and the muon coupling to the percent level. Unless a significantly higher luminosity can be achieved, all other measurements seem to be better done with a hadron or $e^+e^-$ collider. Studies are going on to increase the luminosity at the $\mu^+\mu^- \rightarrow H(126)$ resonance to $10^{32}$ $cm^{-2}s^{-1}$. At higher energies, the muon collider retains its potential as a unique tool to study, via the s-channel resonance, the line shape of possible heavier neutral Higgs bosons, or to investigate very high energy lepton collisions.

## 2.5 Physics of $\gamma\gamma \rightarrow$ Higgs

The photon collider can be seen as an add-on to a linear collider [19] or as a dedicated machine [20]. The reaction of interest is the direct s-channel $\gamma\gamma \rightarrow H$ which has a cross-section of 200 fb. The Higgs cross-section is enhanced for photons of circular polarization in the J=0 state, so the use of a polarized laser allows a significant enhancement of signal over the background. The use of linearly polarized photons allows selection of specific CP states.

The unique attribute of the photon collider is the initial coupling to a pair of photons. Higgs can be observed in e.g. the $\gamma\gamma$ or $b\bar{b}$ final state, following which, using the $b\bar{b}$ partial width measured at another machine, the $H \rightarrow \gamma\gamma$ partial width can be extracted in absolute terms to a precision of 1%. This quantity is of particular interest because this decay proceeds through an inclusive loop that can potentially reveal heavier particles into which the Higgs cannot decay directly. Experimentation would have to be developed to control the luminosity spectrum at the appropriate level of precision.

## 2.6 Higgs Physics Summary

The main purpose of the workshop was to initiate an extended accelerator discussion. It is clear that the physics discussion was only a start, and will need to be continued in a dedicated, broader framework. However a number of important results on the Higgs couplings were already presented.

The relative precisions on Higgs couplings to various particles that were presented for the various proposals are listed in Table 2.1. The table is not yet fully complete but a few conclusions can already be drawn.

- HL-LHC will already be a Higgs factory, able to perform precise measurements on the relative values of the $\gamma\gamma$, gluon-gluon, $t\bar{t}$, W and Z couplings.
- An $e^+e^-$ Higgs factory operating at ZH maximum with the anticipated luminosity for the ILC or LEP3 will access the Higgs boson physics observables that will not be accessible at the LHC (total width, invisible width, and ccbar decay), and allow some improvements over the precision available at the HL-LHC for a few of the other couplings (see Table 2.1), in particular the $b\bar{b}$ coupling.
- To attain the sub-percent precision measurements sensitive to new physics at the TeV scale it is of interest to pursue investigation of accelerators that could give significantly higher luminosities at ZH threshold and below. These are the proposed large circular colliders such as CHF or TLEP.
- Unfortunately, none of the proposed facilities is able to make a very significant measurement of the Higgs self-coupling[1]; investigation of this important question may have to wait for a higher energy collider beyond the LHC/ILC/CLIC.

It should also be emphasized that testing the closure of the Standard Model by precision measurements at the Z peak or the W threshold is one of the important tasks for a next generation lepton collider.

[1] The most significant prospects come from the CLIC studies. Recent and preliminary indications are that a precision of 22% at 1.4 TeV or 11% at 3 TeV could be attained with 1.5 $ab^{-1}$ (resp 2 $ab^{-1}$) integrated luminosity [12].

**Table 2.1:** Expected performance on the Higgs boson couplings from the LHC and $e^+e^-$ colliders, as compiled from the Higgs Factory 2012 workshop. Many studies are quite recent and still ongoing.

| Accelerator → <br> Physical Quantity ↓ | **LHC** <br> 300 $fb^{-1}$ /expt | **HL-LHC** <br> 3000 $fb^{-1}$ /expt | **ILC** <br> 250 GeV <br> 250 $fb^{-1}$ <br> 5 yrs | **Full ILC** <br> 250+350+ <br> 1000 GeV <br> 5yrs each | **CLIC** <br> 350 GeV (500 $fb^{-1}$) <br> 1.4 TeV (1.5 $ab^{-1}$) <br> 5 yrs each | **LEP3, 4 IP** <br> 240 GeV <br> 2 $ab^{-1}$ (*) <br> 5 yrs | **TLEP, 4 IP** <br> 240 GeV <br> 10 $ab^{-1}$ 5 yrs (*) <br> 350 GeV <br> 1.4 $ab^{-1}$ 5 yrs (*) |
|---|---|---|---|---|---|---|---|
| $N_H$ | $1.7 \times 10^7$ | $1.7 \times 10^8$ | $6 \times 10^4$ ZH | $10^5$ ZH <br> $1.4 \times 10^5$ Hνν | $7.5 \times 10^4$ ZH <br> $4.7 \times 10^5$ Hνν | $4 \times 10^5$ ZH | $2 \times 10^6$ ZH <br> $3.5 \times 10^4$ Hνν |
| $m_H$ (MeV) | 100 | 50 | 35 | 35 | 100 | 26 | 7 |
| $\Delta\Gamma_H / \Gamma_H$ | -- | -- | 10% | 3% | ongoing | 4% | 1.3% |
| $\Delta\Gamma_{inv} / \Gamma_H$ | Indirect <br> (30%? ) | Indirect <br> (10% ?) | 1.5% | 1.0% | ongoing | 0.35% | 0.15% |
| $\Delta g_{H\gamma\gamma} / g_{H\gamma\gamma}$ | 6.5 – 5.1% | 5.4 – 1.5% | -- | 5% | ongoing | 3.4% | 1.4% |
| $\Delta g_{Hgg} / g_{Hgg}$ | 11 – 5.7% | 7.5 – 2.7% | 4.5% | 2.5% | < 3% | 2.2% | 0.7% |
| $\Delta g_{Hww} / g_{Hww}$ | 5.7 – 2.7% | 4.5 – 1.0% | 4.3% | 1% | ~1% | 1.5% | 0.25% |
| $\Delta g_{HZZ} / g_{HZZ}$ | 5.7 – 2.7% | 4.5 – 1.0% | 1.3% | 1.5% | ~1% | 0.65% | 0.2% |
| $\Delta g_{HHH} / g_{HHH}$ | -- | < 30% <br> (2 expts) | -- | ~30% | ~22% <br> (~11% at 3 TeV) | -- | -- |
| $\Delta g_{H\mu\mu} / g_{H\mu\mu}$ | < 30% | < 10% | -- | -- | 10% | 14% | 7% |
| $\Delta g_{H\tau\tau} / g_{H\tau\tau}$ | 8.5 – 5.1% | 5.4 – 2.0% | 3.5% | 2.5% | ≤ 3% | 1.5% | 0.4% |
| $\Delta g_{Hcc} / g_{Hcc}$ | -- | -- | 3.7% | 2% | 2% | 2.0% | 0.65% |
| $\Delta g_{Hbb} / g_{Hbb}$ | 15 – 6.9% | 11 —2.7% | 1.4% | 1% | 1% | 0.7% | 0.22% |
| $\Delta g_{Htt} / g_{Htt}$ | 14 – 8.7% | 8.0 – 3.9% | -- | 5% | 3% | -- | 30% |

(*) The total luminosity is the sum of the integrated luminosity at four IPs.

Comments on Table 2.1:

- For the LHC:
  Measurements at the LHC are extracted from final state Higgs decay cross-sections. Since there is no tagged Higgs channel at the LHC there is an unknown overall scaling factor that can be taken as either an unknown total width normalization or an uncertainty in invisible channels. Except for invisible width limits, the precisions given in the table were obtained under the assumption that there is no invisible decay mode of the Higgs boson so that the total width is the sum of the observed partial ones. One can alternatively, and without model dependence, interpret these numbers as precision of the relative couplings. The first set of numbers corresponds to the hypothesis in which the systematic errors remain the same as in today's LHC results; the second set of numbers corresponds to the assumption that experimental systematic errors scale down with statistics, while the systematic errors of theoretical nature (flux factors) could be reduced by a further factor of two – this second assumption is considered the most realistic at the moment, but does not take into account possible improvement in the detectors. Except for $g_{HHH}$, the estimated precisions are for one single LHC experiment.
- For the the $e^+e^-$ facilities – The quoted integrated luminosities and the expected sensitivities are based on the following assumptions:
  (a) All luminosity numbers were taken at face value from the proponents of the various facilities (Tables 8.1 and 8.2). Some are optimistic, others conservative. The integrated luminosities correspond to the total delivered by each facility. It is the sum of two experiments sharing one IP at ILC, or the sum over four experiments for LEP3 and TLEP. Operating time of $10^7$ seconds per year was assumed.
  (b) The integrated luminosities for the ILC and CLIC were based on a model with slow initial build-up for machine operation.
  (c) The numbers of Higgs for the CLIC do not include the effects of beam polarization except for the HHH coupling studies.

# 3 Linear $e^+e^-$ Colliders

## 3.1 Introduction

ILC R&D has been based on a truly global international collaboration. In its RDR and TDR phases, the Global Design Effort (GDE) has been the global working force on ILC accelerator R&D with about 130 participating institutions (http://www.linearcollider.org). The core technology for ILC of 2×10 km linacs is Superconducting RF (SRF) technology. It features seventeen thousand 1.3 GHz SRF cavities with accelerating field of 31.5 MV/m in 1,700 cryomodules and is based on over 20 years worldwide R&D efforts. The TDR was completed at the end of 2012. The volume covering accelerators consists of two parts: Part I: R&D, Part II: Baseline Reference Report. Based on the TDR, staging scenarios to start with a Higgs factory of 250 GeV extendable to 1 TeV are proposed.

CLIC is an international collaboration of 44 institutions from 22 countries. CLIC has a staged design to reach a center-of-mass energy of 3 TeV. Different from conventional klystron powered linac, CLIC adopts a drive beam scheme to produce the main linac RF (distributed klystron scheme), with a drive beam current ~100 times greater than the accelerated beam current. In 2012, CLIC has published a CDR in three volumes: Vol. I: The CLIC accelerator and site facilities, Vol. II: Physics and detectors at CLIC, Vol. III: CLIC study summary. The feasibility of the CLIC scheme has been established. The CLIC staging scenarios include a first stage at 500GeV that can be used as a Higgs factory. Higher energy stages will still contribute to the Higgs studies, e.g. to the measurement of the Higgs self-coupling and the coupling to WW.

The main advantage of the linear collider is that the cost grows as $A+B\times E_{CM}$, thus "only" linearly with energy; the main drawback is that both A and B are very large. The challenge is to achieve reliably very high bunch intensities and very small beam sizes for the two beams simultaneously and to collide them. There are specific proposals (ILC and CLIC) which have been designed to the point that the main technologies are in hand, although the level of readiness are somewhat different. International collaborations have been set-up and organized. The luminosity grows linearly with energy with a value of $2\times10^{34}$ $cm^{-2}s^{-1}$ at $E_{CM}$ = 500 GeV. The luminosity can be doubled by increasing the RF power by a factor of about 1.5 at a capital cost increase of several percent. The machine delivers beam to one IP. High level (80%) of any type of beam polarization is readily available for electrons, whereas a lower level (30%) is achievable for positrons. Beam energy calibration relies on accurate beam spectrometers with a relative precision of a few $10^{-4}$. Beamstrahlung induces a broadening and some uncertainty in the center-of-mass energy, but this is not of great importance for Higgs physics. The collision environment features electromagnetic residues which have been carefully studied. The beam comprises 5 bunch trains per second which allows the use of very thin detectors operating in a pulsed mode. There exists a proposal to establish a linear collider in Japan starting with a center-of-mass energy of 250 GeV.

Compared with other types of Higgs factory, linear $e^+e^-$ Higgs factories have the following features:

- Advantages:
  - Extensive design and prototyping work have been done.
  - Key technologies are in hand after large investment for R&D.
  - There exist well-organized international collaborations led respectively by the ILC GDE and CLIC Collaboration (soon to be combined in the Linear Collider Collaboration).
  - It is an important step towards high energy e+e- collisions.
  - Polarized beams ($e^-$ 80%, $e^+$ 30%) can be created.
  - It is the front runner (in terms of readiness).
- Challenges:
  - High cost
- Specific issues:
  - ILC
    - Final Focusing System (FFS)
    - Positron source for a Higgs factory needs 10 Hz operation of the electron linac for e+ production, or the use of an unpolarized e+ beam as a backup scheme

- CLIC
    - Accelerating structure
    - Industrialization of major components
    - From CDR to TDR

The KEK x-band $e^+e^-$ / $\gamma\gamma$ Higgs Factory is based on CLIC-type cavities and the existing conventional RF technology. An optical FEL can also be used to produce high-energy photon beams for a $\gamma\gamma$ collider. In the first stage of operation ($e^+e^- \rightarrow$ Z, WW and $\gamma\gamma \rightarrow$ H) the proposed facility could be built on the KEK site. If the initial operational mode is with photon beams, then there is no need for an e+ source or a positron damping ring. The total length of the two linacs would be short, about 3.6 km at $E_{CM}$(ee) ~ 250 GeV.

The machine-detector interface (MDI) for ILC 500 GeV and CLIC 3 TeV has been studied. The requirements for MDI are to provide reliable collisions of ultra-small beams (~few nanometers), with an acceptable level of background.

## 3.2 ILC-based Higgs Factory

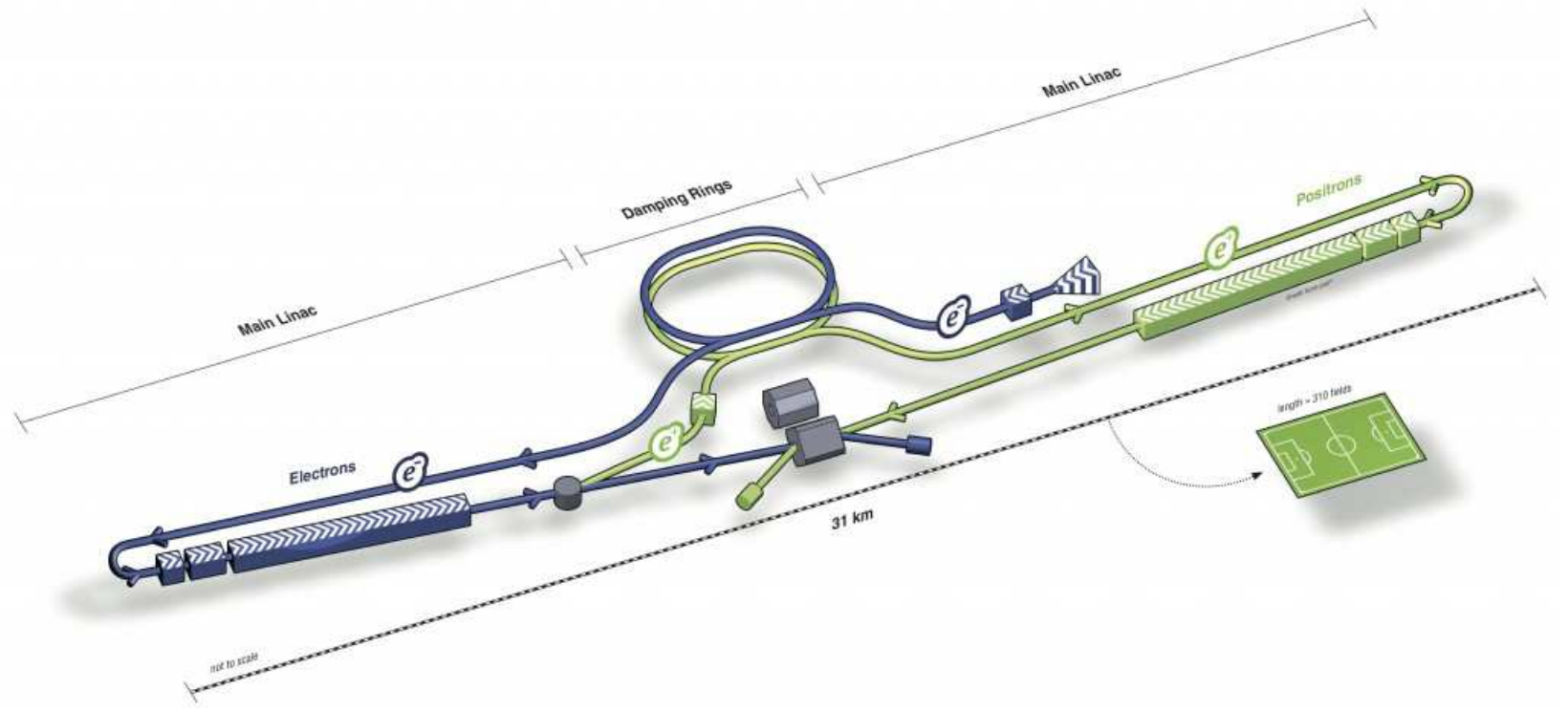

**Figure 3.1:** Layout of the ILC.

ILC is in the advanced design stage. (Figure 3.1) The Technical Design Report (TDR) was completed at the end of 2012. The official printed version will appear in June in 2013 after a few review steps. The TDR concentrates on the design of a machine of center-of-mass energy $E_{CM}$ = 500 GeV and does not describe in detail the design as a Higgs Factory. However, the technology for the Higgs Factory is obvious. Up to the center-of-mass energy 500 GeV ILC can adapt to any staging scenarios required by physics. The difference is only the lengths of the linac and the tunnel. The estimated cost of the 250 GeV Higgs Factory is 67% of the cost of the 500 GeV collider and is 75% if the tunnel for 500 GeV is constructed. The required total site power is about 120 MW and 125 MW, respectively. (The power for 500 GeV operation is 160 MW.) In the present scope of ILC, when going beyond $E_{CM}$ = 500 GeV, an R&D is planned for higher accelerating gradients for cost savings. Even without higher gradient the cost increase for the 1 TeV machine is around 10% of the 500 GeV machine.

In the case of $E_{CM}$ < 300 GeV, 10Hz operation (5 Hz for collision and 5 Hz for positron production) is planned in the baseline design because the positron production in the undulator scheme is somewhat inefficient when the electron energy $E_{CM}$ /2 is lower than 150 GeV. (Figure 3.2) This does not cause any problem technically, but is not very elegant. The 10 Hz operation requires an extra length of electron linac corresponding to 150 – 250/2 = 25 GeV and extra electric power ~25 MW. Thus, if 10 Hz operation is avoided, the construction cost ~3% and the operation power ~25 MW can be saved compared with the values quoted in the TDR. A possible way to avoid 10 Hz operation is to adopt the electron-driven, conventional method. This is a proven design. The only change in physics is that the positron is unpolarized.

In all cases the luminosity presented in Table 8.1 assumes 1312 bunches per pulse. This can be upgraded by a factor of 2 by adding about 50% more RF system.

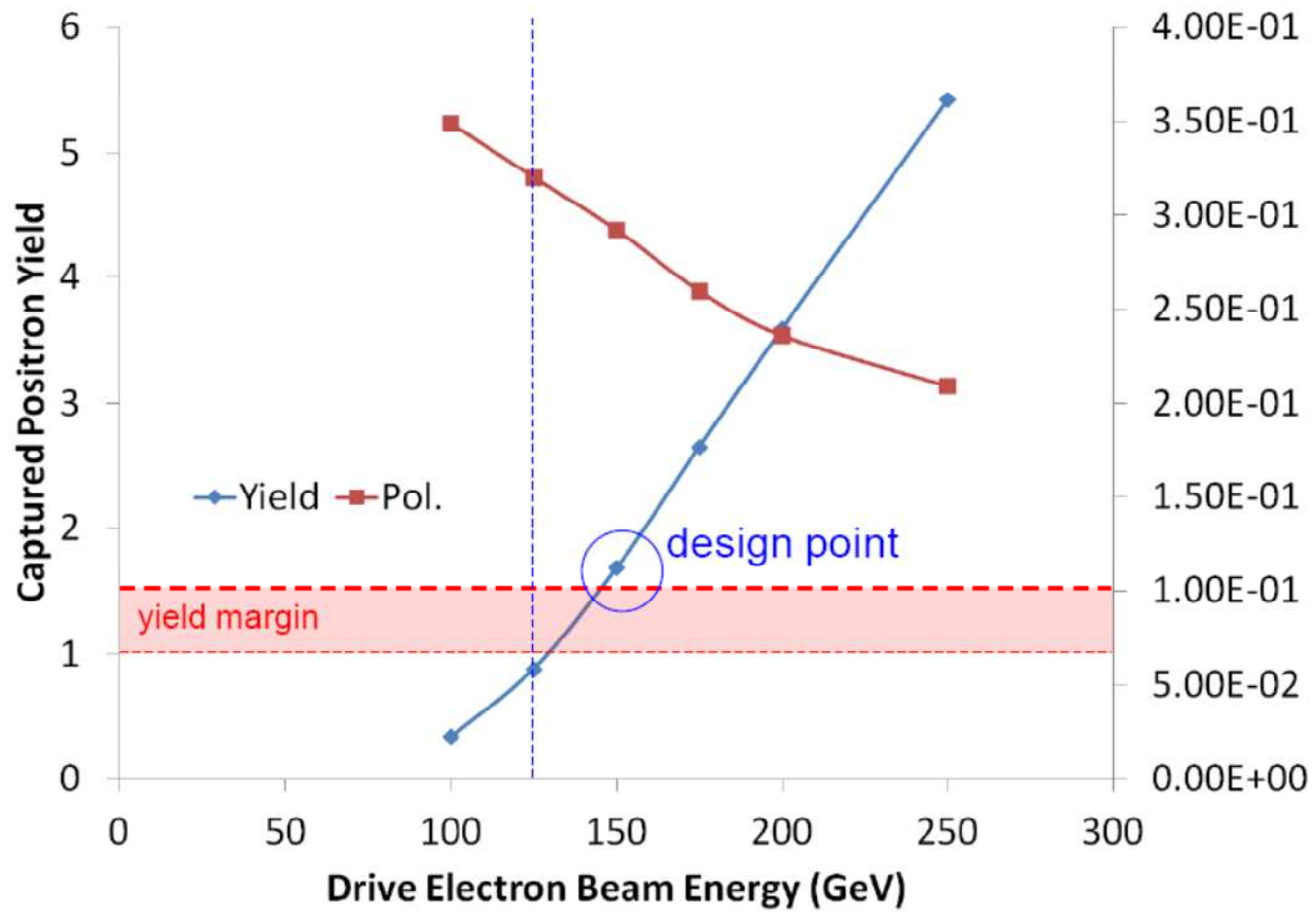

**Figure 3.2:** Positron yield (blue) and polarization (red) as a function of the drive electron beam energy.

There are a few items that require final steps of R&D.

First, the target for the positron production still needs several months or 1-2 years of further study. However, the design of the backup scheme, which uses conventional electron-driven system, has been completed. If ILC has to start construction in 2013, this backup scheme can be adopted. The only disadvantage is that positron polarization would not be available.

Second, the test of the final focusing system is still going on at KEK-ATF2. There have been delays, including that due to the 2011 earthquake, but no fundamental problem is expected. The latest operation in December 2012 achieved a vertical beam size ~70 nm (Figure 3.3), which differs from the design (37 nm) by only a factor of two. Further study is being planned in 2013. [21]

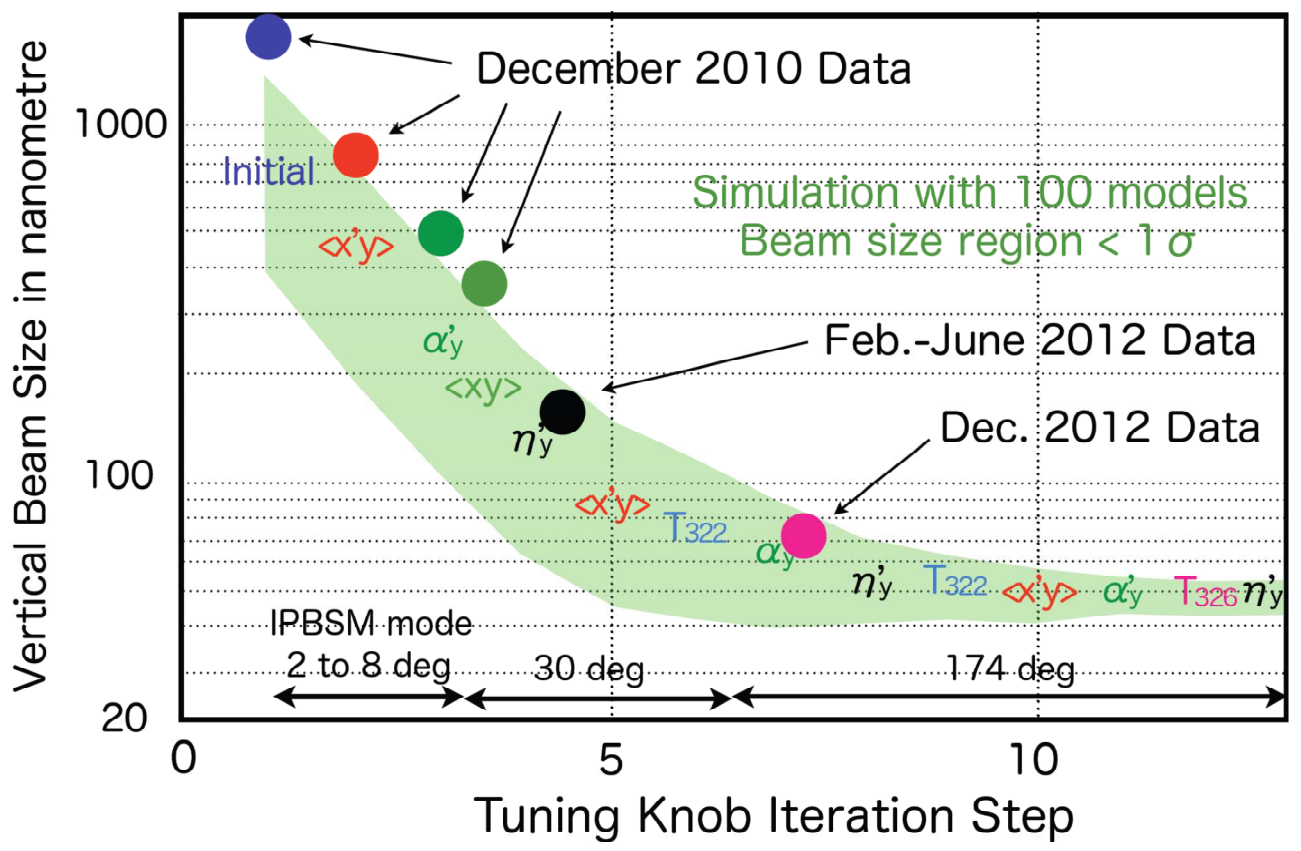

**Figure 3.3:** Vertical beam size from the ATF2 experiment.

## 3.3 CLIC-based Higgs Factory

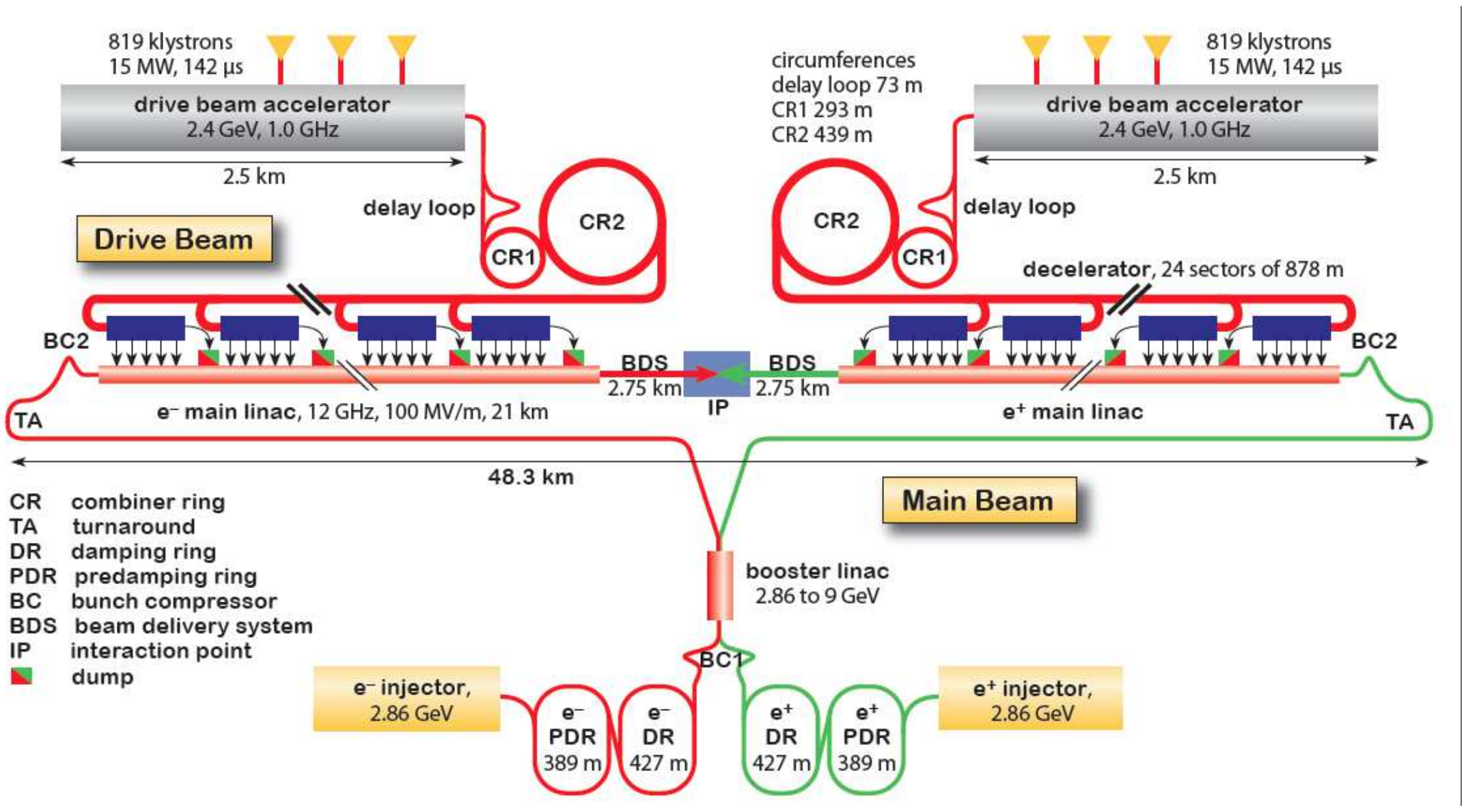

**Figure 3.4:** Layout of the CLIC.

Figure 3.4 shows the layout of the CLIC. The CLIC baseline design proposes building in stages of increasing energy. The choice of these stages will have to be fixed taking into account future LHC results. Currently two example scenarios exist, named A and B, in order to illustrate the staging strategy. Scenario A has stages at 500 GeV, 1.4 TeV and 3 TeV, while scenario B has stages of 500 GeV, 1.5 TeV and 3 TeV. Scenario A provides higher luminosity at 500 GeV, while scenario B has a lower integrated cost for the three stages. Each of the energy stages contributes to the Higgs studies. Table 3.1 lists the parameters of the 500 GeV and 3 TeV stages of scenario A.

It is possible to operate each energy stage of CLIC at a lower than nominal energy. This can be achieved by operating the main linac at a reduced gradient, i.e. by reducing the drive beam current. The main bunch charge also needs to be reduced in this mode in order to preserve the same beam quality. At certain reductions of the gradient it is possible to increase the length of the drive and main beam pulses. The resulting increased number of bunches per beam pulse allows an increase in luminosity. The main and drive beam complex is fully prepared for this type of operation. The power consumption at lower than nominal collision energy is somewhat lower than at 500 GeV, the exact value depends on the collision energy.

The CLIC physics study group envisages performing the Higgs measurements at a center-of-mass energy of 500 GeV or 350 GeV, rather than at 250 GeV. The total number of Higgs that can be expected at different center-of-mass energies is shown in Figure 3.5 for the $e^+e^- \rightarrow ZH$ and $e^+e^- \rightarrow \nu\nu H$; the parameters for 250 GeV and 350 GeV are given in Table 3.1.

If large importance was attributed to the operation at 250 GeV, one could also consider adding an extraction line in the main linac, to extract the beam at 125 GeV. This would result in somewhat increased luminosity. The same luminosity would be obtained if CLIC were built for this collision energy only. However this is currently not foreseen since it delays the program at higher energies.

The feasibility of the CLIC concept has been established with the studies documented in the CDR. In particular the drive beam concept has been proven at a test facility (CTF3) and very high gradients have been achieved experimentally. Specific challenges as the unprecedented alignment and stability tolerances have been successfully addressed experimentally. The main remaining challenges are to develop a technical design, based on the conceptual design. This includes an optimization of the accelerator components and systems and preparation for industrial procurement. In particular small series production of accelerating structures is important.

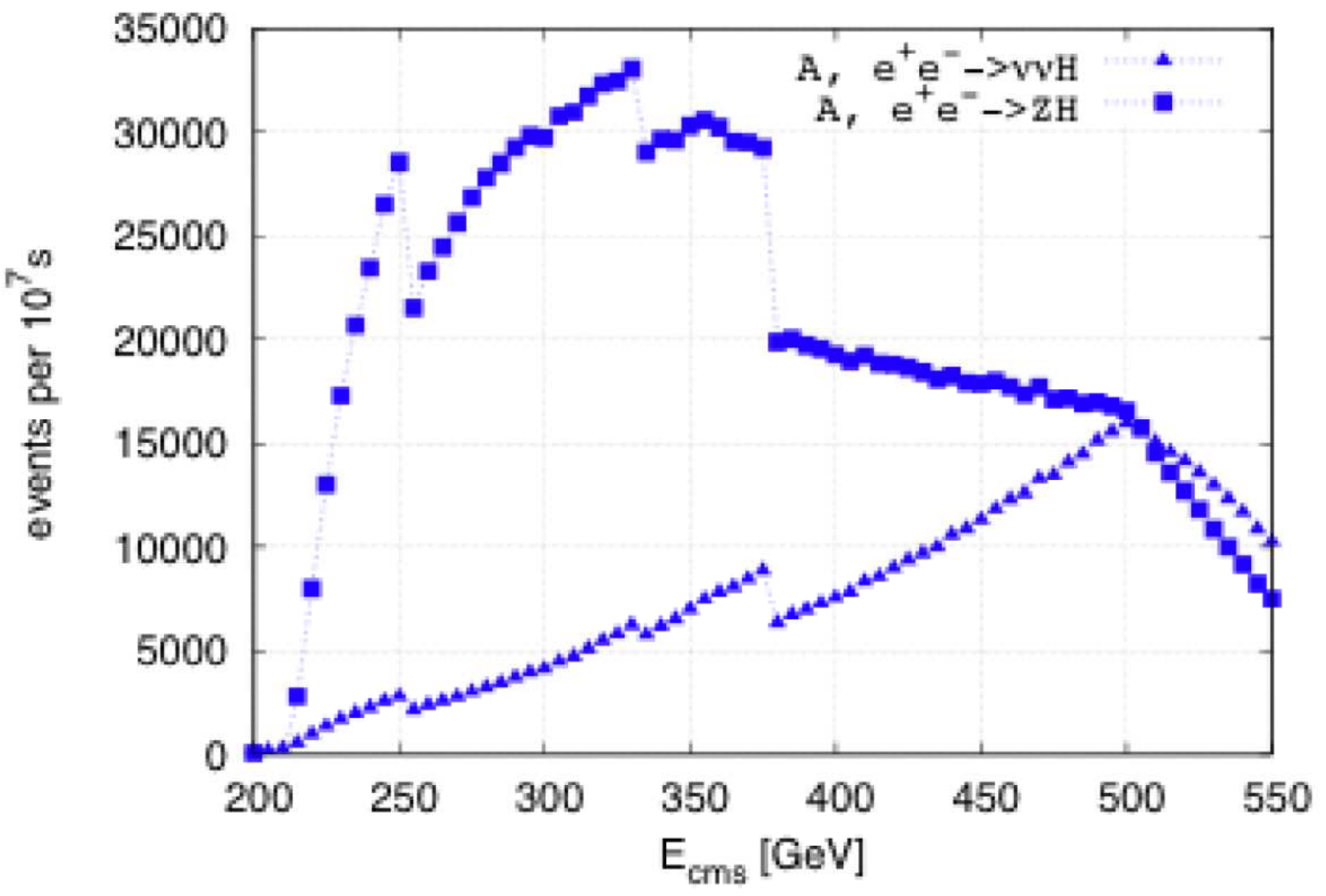

**Figure 3.5:** The rate of Higgs bosons produced in the 500 GeV stage of CLIC scenario A operated at different center-of-mass energies. The design is slightly modified with respect to the CDR: the same number of wigglers has been installed in the damping ring as for the 3 TeV case. The quality of the luminosity spectrum is in all cases similar or better than at 500 GeV.

**Table 3.1:** The luminosity at different energies for scenario A. The effective cross section for Higgs production and the number of Higgs per $10^7$ sec is also given. It has been calculated based on a parameterization of the cross section derived with Wizzard2 [22] and adding the beam-beam and initial state radiation with GUINEA-PIG. Polarisation has been neglected, which increases the rate for $e^+e^- \rightarrow \nu\nu H$ by 80%.

| **Design $E_{CM}$** [GeV] | **500** | | | **3000** |
|---|---|---|---|---|
| **Operating at $E_{CM}$** [GeV] | **250** | **350** | **500** | **3000** |
| L [$10^{34}$ cm$^{-2}$s$^{-1}$] | 1.37 | 2.13 | 2.3 | 5.9 |
| $L_{0.01}$ [$10^{34}$ cm$^{-2}$s$^{-1}$] | 1.04 | 1.30 | 1.4 | 2.0 |
| σ ($e^+e^- \rightarrow \nu\nu H$) [fb] | 20.4 | 32.4 | 67.6 | 415 |
| σ ($e^+e^- \rightarrow ZH$) ) [fb] | 208.2 | 141.3 | 70.1 | 4.6 |
| $e^+e^- \rightarrow \nu\nu H$ per$10^7$ sec | 2,795 | 6,901 | 15,548 | 244,850 |
| $e^+e^- \rightarrow ZH$ per$10^7$ sec | 28,551 | 30,097 | 16,123 | 2,714 |

## 3.4 X-band Klystron-based Higgs Factory

An X-band $e^+e^-$ / $\gamma\gamma$ Higgs factory studied at KEK is a conventional klystron-based facility. Linear accelerators are based on CLIC-type cavities and the existing RF technology (XL4 klystrons, ScandiNova modulators, SLED II system). A two-beam scheme could be implemented at a later stage as a test facility for CLIC. An optical FEL can be used to produce high-energy photon beams for a $\gamma\gamma$ collider. In the first stage of operation ($e^+e^- \rightarrow Z$, WW and $\gamma\gamma \rightarrow H$), the proposed facility can be built on the KEK site. If the initial operational mode is with photon beams, then there is no need for an e+ source or a positron damping ring. With a crossing angle of ~25 mrad for both $e^+e^-$ and $\gamma\gamma$ beams, only a single set of beam dump lines would be required. For some processes, the required center-of-mass energy is considerably lower at the proposed facility than at an $e^+e^-$ collider. The rich set of final states in $e^+e^-$ and $\gamma\gamma$ collisions is instrumental for measuring the properties of the Higgs boson.

## 3.5 Machine-Detector Interface

The Machine Detector Interface (MDI) for linear $e^+e^-$ colliders, such as ILC and CLIC, allows for essentially full solid angle acceptance. Beam parameter measurements (energy, polarization, luminosity) allow control of beam-beam effects on physics analyses. MDI magnet technologies are well into development, such as a compact SC magnet for the ILC, and a hybrid permanent magnet for the CLIC. The risks to machine performance due to single pass collisions of nm-size bunches will be controlled via IP luminosity feedback, especially advantageous is IP feedback for the ILC bunch structure.

# 4 Circular $e^+e^-$ Colliders

## 4.1 Introduction

A number of proposals exist for Higgs factories based on a circular $e^+e^-$ collider with a center-of-mass energy of 240 GeV and in some cases extending to 350 GeV or 500 GeV. Circular colliders have been successfully used for lower energy machines in the past and the concept and technology are well developed. The highest energy facility was LEP2, which reached a maximum center-of-mass energy of 209 GeV. Circular colliders allow for more than one interaction point and potentially could provide more luminosity than linear colliders of equal energy. However their energy reach is limited and a number of issues may compromise their performance as will be detailed below. The proposals are in a very early stage with no design reports, which makes it difficult to evaluate them. A brief summary of the advantages and challenges of circular $e^+e^-$ colliders in general are given below:

- Advantages:
    - At 240 GeV and below, a higher luminosity than a linear collider when the ring is sufficiently large

- ➢ Based on mature technology and rich experience
  - ➢ Some designs can use existing tunnel and site
  - ➢ More than one IP
  - ➢ Tunnel of a large ring can be reused as a *pp* collider in the future
- Challenges:
  - ➢ Beamstrahlung limiting beam lifetime and requiring lattice with large momentum acceptance
  - ➢ RF and vacuum problems from synchrotron radiation
  - ➢ A lattice with low emittance
  - ➢ Efficiency of converting wall power to synchrotron radiation power
  - ➢ Limited energy reach
  - ➢ No comprehensive study; design study report needed.

## 4.2 Circular $e^+e^-$ Colliders Considered

The main parameters for the different proposals considered at the Workshop are listed in Table 8.2.

The driving parameter of a circular collider accelerator design is its circumference, which is in some cases determined by external constraints and in some cases by a cost optimization. Otherwise the design strategy is fairly similar for all proposals. Based on the choice for the ring circumference, the colliders can be grouped as follows:

- LEP3 has a circumference of 26.7 km in order to be installed in the existing LHC tunnel, leading to a serious cost reduction. However, installation of LEP3 for concurrent operation with the LHC is complicated and unlikely.

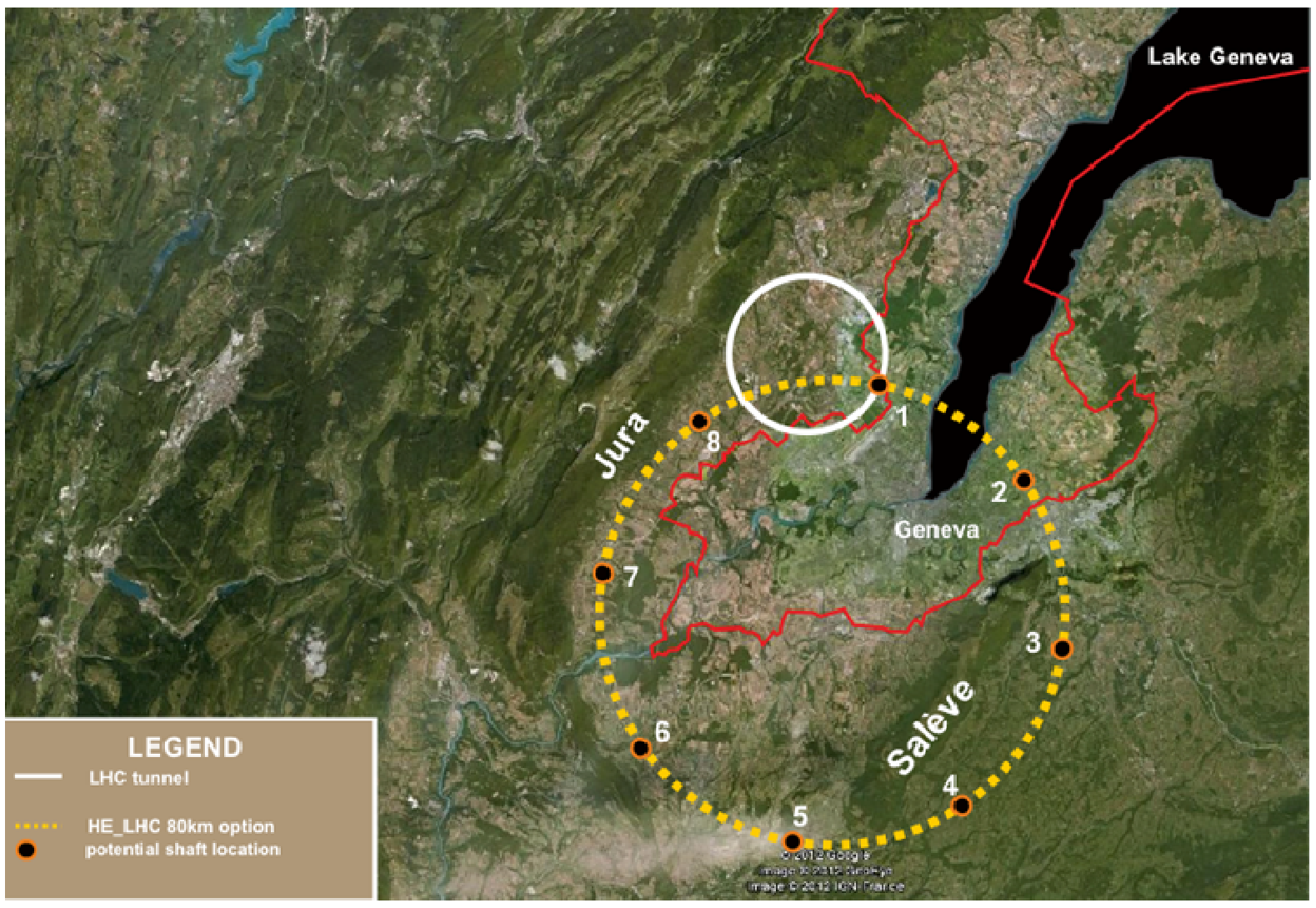

**Figure 4.1:** Sketch of LEP3 (white circle) and TLEP (yellow circle).

- As its name suggests, the Fermilab site-filler would fit on the Fermilab site. It therefore is limited to a smaller circumference than the other designs of about 16 km. It is conceivable to later reuse this machine as an injector for a very large hadron or lepton collider with a very large circumference in excess of 200 km.

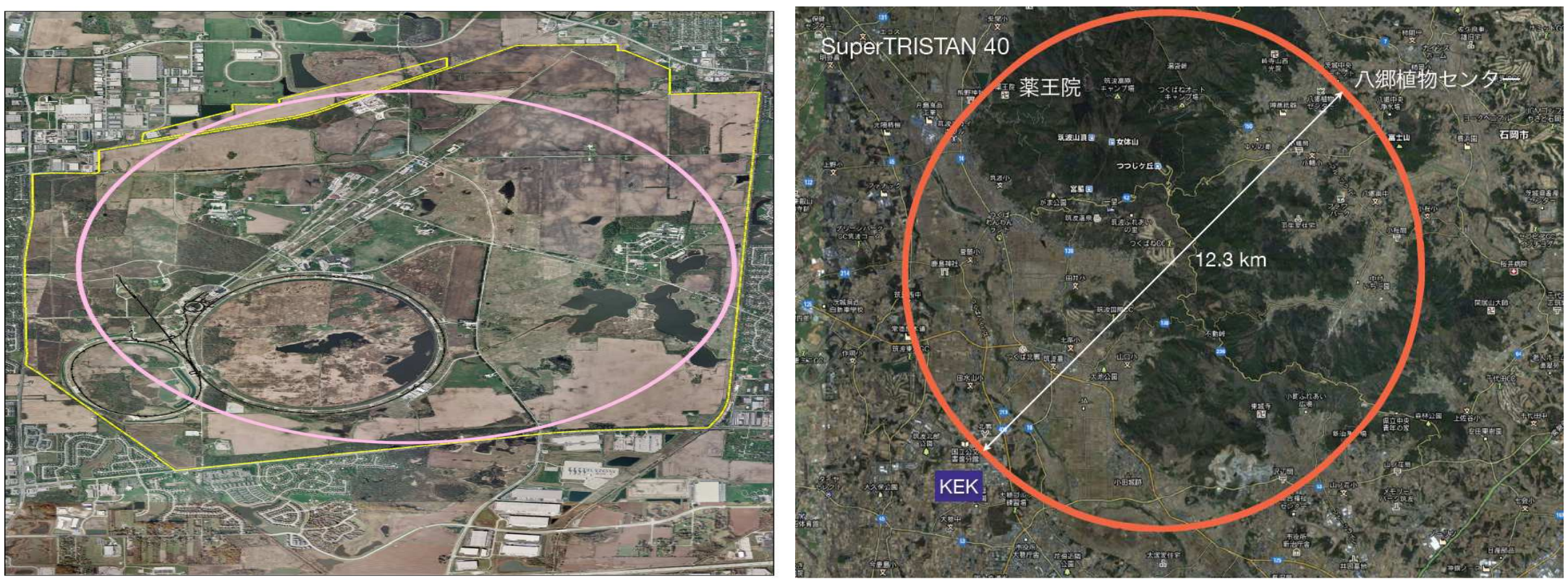

**Figure 4.2:** Left: Fermilab site-filler; Right: SuperTRISTAN.

- TLEP, the different versions of SuperTRISTAN and the IHEP Higgs Factory as well as VLLC use larger circumferences on the order of 40 to 233 km with the aim to optimize the machine performance and to be able to reach 350 GeV and in the case of VLLC with 233 km circumference also 500 GeV. In these cases one can conceive of installing a hadron collider in the same tunnel at a later stage. This collider could have proton energies a factor of a few higher than LHC, the exact value depending on the size of the ring and the magnet technology used. One could even contemplate lepton-hadron collisions.

## 4.3 Technical Challenges

The main challenges of the various proposals are fairly similar. Basically, the storage ring collider technology is well established, so the technical issues have been based on well-established accelerator physics and technologies of the past including the LEP2, the B-factories and the circular synchrotron light sources. A possibility that some unforeseen technical issues are waiting around the corner, although present, is relatively small. Some of the issues below, however, are critical because they require various degrees of extrapolations from past experience, and because at this time there has been very little conceptual design or R&D work devoted to these circular colliders.

### 4.3.1 Energy reach and upgradability

The proposed center-of-mass energy is 240 GeV for all machines. For the larger rings TLEP and SuperTRISTAN also 350 GeV is proposed. It is considered that reaching higher energies further would lead to a strong increase in the cost of the projects and is therefore not practical or even forbidding. In particular the potential to increase the energy of an existing circular collider will be very limited. Operation at the Z peak and W pair threshold can be envisaged with luminosities 2-5 orders of magnitude higher than in the LEP.

The energy that can be reached with a circular $e^+e^-$ collider is determined by the size of the ring and the installed RF voltage, which are both important cost factors. The circulating beams emit synchrotron radiation. The average energy loss $E_{loss}$ of each particle per turn is given by $E_{loss}$ = 88 keV $(E/GeV)^4/(\rho/m)$, for a beam energy E and bending radius ρ. For example, a 120 GeV electron in the LEP/LHC tunnel would emit about 7 GeV per turn. This loss needs to be compensated with accelerating RF with a total voltage exceeding the loss. In order to reach a higher energy, either the RF voltage or the radius, or both, has to be increased. A simple cost model can be applied to conclude that the RF voltage and the cost of accelerator are expected to scale as $E^2$. Circular colliders are therefore basically machines operating between the Z peak and the 240 GeV ZH cross-section maximum, possibly up to the top threshold at 350 GeV. They become highly unpractical approaching the level of 500 GeV, and impossible substantially above 500 GeV. The decision on whether or not to go in the direction of circular colliders depends critically on what required center-of-mass energy is needed to explore the Higgs coupling constants. Without this critical input, no useful decision can be made.

For the larger rings, one can however reuse the tunnel to install a hadron machine later, in a similar fashion as the LHC has been installed in the LEP tunnel. This would provide a path to future projects. For the Fermilab site-filler one can conceive using it as an injector for a larger ring. LEP3 does not provide an upgrade path but rather exploits existing infrastructure.

### 4.3.2 **Synchrotron radiation**

The luminosity of a circular collider increases with the circulating beam current, which in all proposals is several mA. The current is limited in all proposals by the total power of the emitted synchrotron radiation, which is typically set to about 100 MW (LEP2 went up to 20 MW). The main challenges are to limit the power consumption by obtaining good efficiency for the transfer from wall plug power into beam power and to deal with the impact of the intense radiation on the vacuum and the RF. In addition, power requirements other than compensation for synchrotron radiation, such as for cryogenics, magnets, water cooling, injectors etc., need to be taken into consideration and optimized.

Compared to LEP2 the synchrotron radiation is more intense and in some designs also the critical energy is higher, which might cause a significant radiation hazard and this issue needs to be evaluated and solved.

A larger ring circumference obviously allows both increasing the beam current linearly with the radius for the same radiation power and reducing the installed RF voltage inversely with the radius. As a result the luminosity will increase linearly with the bending radius. With beam lifetime included (see section 4.3.6 Beamstrahlung), the luminosity gains with the bending radius even more rapidly than linearly.

### 4.3.3 **RF system**

The RF system must provide the high accelerating voltage to overcome the synchrotron radiation loss with sufficient overvoltage to provide an acceptable quantum lifetime. It must also provide the high power to be transferred with high efficiency to the beam. The RF system must be shielded against the strong synchrotron radiation.

Distribution of the RF stations requires a conceptual design to minimize the orbital saw-tooth effects, especially in the cases when $e^+$ and $e^-$ beams occupy the same vacuum chamber.

An optimal choice of the RF frequency needs to be made. An efficient RF power coupler for the needed system also needs to be worked out in a conceptual design of the collider. Note that the ILC cavities (1.3 GHz) have not been designed for this high average power and that the coupler R&D would need to be redone. At 700 MHz, however, the work done for the high power proton machines can be readily applied.

When a large ring is operated at a relatively low beam energy E at a fixed synchrotron radiation power, the required beam current can become large and increase as $1/E^4$. (For example in the case when the TLEP is operated at the Z-pole, the beam current exceeds 1 Ampere.) Since the RF section is very long compared with existing rings, possible collective instabilities should be studied. The RF section length can be made shorter at low energies in view of the total voltage required, but the demand to the power coupler would be even more severe when the total synchrotron radiation power is fixed.

### 4.3.4 **Vacuum system**

The raw synchrotron radiation power per meter will be much higher than LEP2 although lower or comparable to that of the B-factories. Although some other parameters will require much extrapolation from existing colliders, most of the vacuum system issues such as the higher-order mode cooling are expected to be within our engineering capacity. The one most significant exception might be the radioactivation by the high critical energy (~1.5 MeV) and high power synchrotron radiation. It is possible that this consideration may drive the entire vacuum chamber design.

### 4.3.5 **Beam-beam effects**

Much of the past experience in operating the LEP and the B-factories has been incorporated into the proposals of the circular Higgs factories. The achieved beam-beam parameter of 0.083 per IP (Figure 4.3) with 4 simultaneously operating IPs at LEP2 [23] provided a solid basis for the design of the new colliders. However, not all past experiences have been consistent and there remains a need to continue the study and identify and understand the underlying beam dynamics. Keeping in mind the critical impact of beam-beam effects on the luminosity, the new operating regime of beamstrahlung and collision point optics, it is still critical to fully understand the beam-beam effects in order to optimize the design. On the other hand, one possibility on the optimistic side could be that the beam-beam parameter value might be substantially increased due to the much enhanced synchrotron radiation damping.

With high RF voltage, it is expected that the synchrotron tune of the electron beam will be high. Together with the small $\beta_y$ at the interaction point, which leads to large hour-glass effect, it may generate potentially harmful beam dynamics effects.

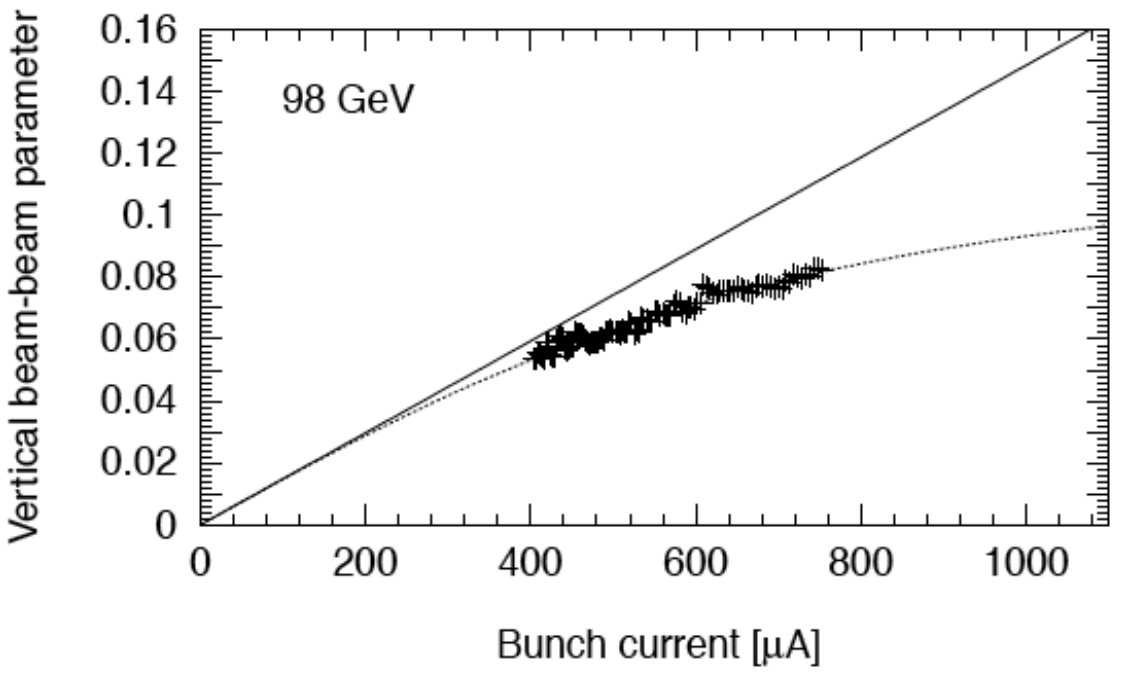

**Figure 4.3:** Vertical beam-beam parameter measured at LEP2.

### 4.3.6 **Beamstrahlung**

The strong beam-beam forces also lead to the emission of energetic photons, i.e. beamstrahlung, which reduces the beam particle energy and leads at each collision to the development of a low energy tail in the beam. (Figure 4.4) [24] The effect is weaker than the beamstrahlung in the linear colliders and does not affect the collision energy spread significantly. However, due to the limited energy acceptance of the machine, particles in this tail can be lost thus reducing the luminosity lifetime. In order to keep the beam longer than a typical refill time of once per minute, less than one particle in ~$10^6$ must be lost per bunch crossing. A lattice and RF energy acceptance of 2-6% is required and the beam parameters need to be adjusted to reduce beamstrahlung; this could potentially result in a reduction of luminosity.

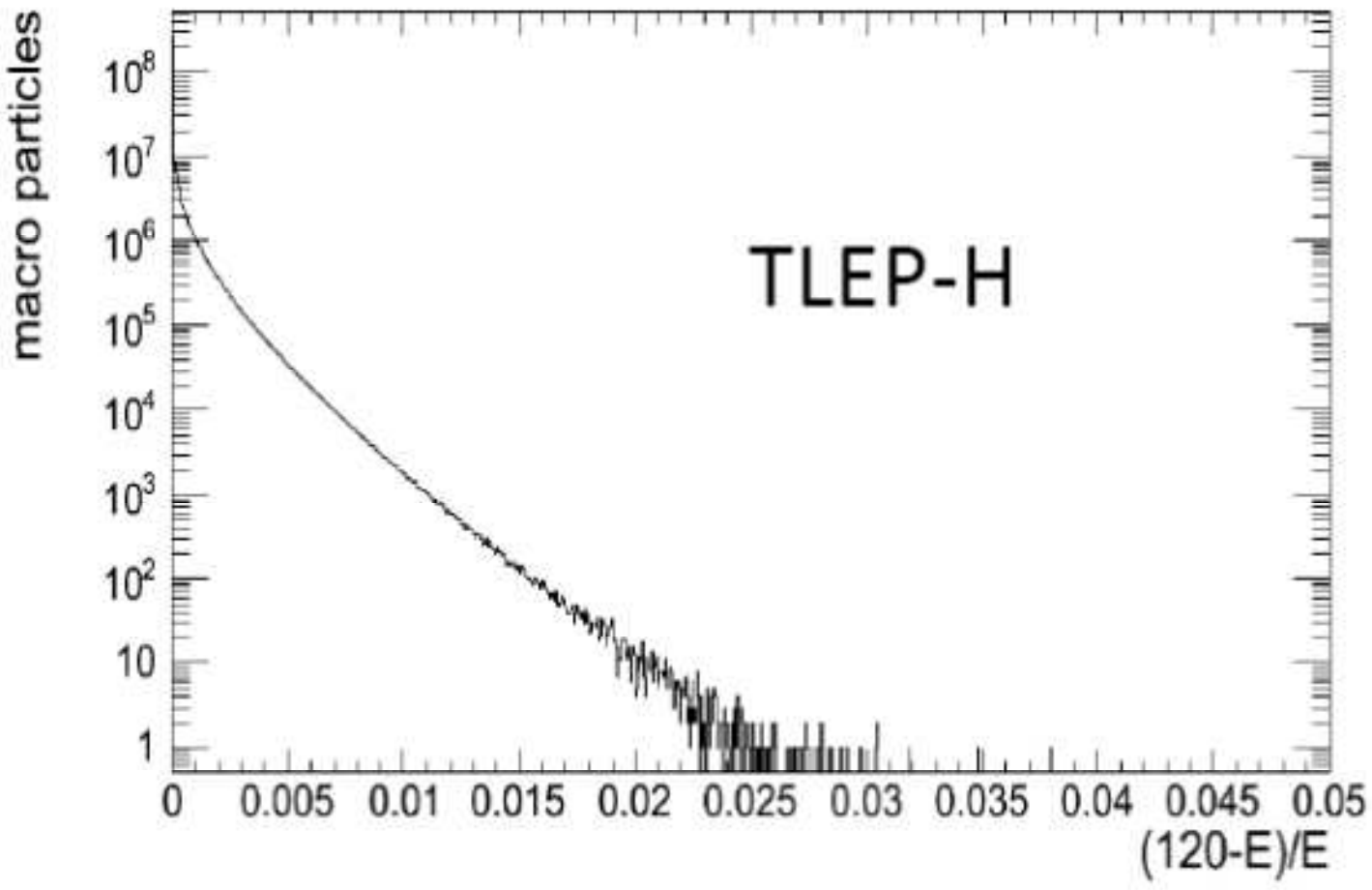

**Figure 4.4:** Simulation of the particle spectrum after one collision due to the beamstrahlung effect at TLEP. An energy acceptance of 2% is necessary.

### 4.3.7 **Lattice design**

The lattice design is challenging, since small beta-functions at the interaction point and a large distance between the interaction point and the first quadrupole magnet are required together with a large energy acceptance to reduce the impact of beamstrahlung on the luminosity lifetime.

Decisions need to be made on whether the collider will consist of a single ring or a double ring. It also requires an early decision on colliding beams head-on or at an angle.

Special and dedicated efforts must be invested in the interaction region design because of its complexity, its direct impact on the optimization of beam-beam effects and luminosity, and its impact on the final energy acceptance. Diagnostics need to be planned carefully in this region.

Also important is the machine-detector interface that impacts on the accelerator performance, the detector performance, and the expected detector noise background level. Due to synchrotron radiation and other backgrounds, the IP vacuum chamber may require a small size that compromises high precision flavor tagging.

A significant development was made during the workshop in terms of an idea to increase the energy acceptance due to the interaction region optics. Further effort to consolidate this idea as well as further optimization will be needed in this important area in order for the circular collider design to be completed.

Lattice design for the regular cells is less of an issue, particularly when considering the experiences gained in low-emittance light sources.

### 4.3.8 **Emittance control**

Various effects can disturb the small beam emittance required for the collider operation. In addition to beamstrahlung, these include dynamic aperture and nonlinear optical effects, intrabeam scattering, electron-cloud, higher-order-mode heating with short bunches, coherent synchrotron radiation instability, etc.

An important issue is to control the ratio of vertical and horizontal emittance in the collider ring. It is necessary that this ratio be minimized before the beam-beam collisions blow up the beam due to beam-beam effects. A large number of error tolerances, as well as diagnostics tools and correction mechanisms, will need to be established to assure a small value for this emittance ratio. Operational experience gained at synchrotron radiation facilities can be very helpful here as well.

### 4.3.9 **Top-up injection**

Due to high luminosity and the beamstrahlung effect, the beam lifetime would be limited. In order to keep the luminosity nearly constant during a production cycle, a second ring as an injector would be added for the top-up injection scheme, which was successfully employed at PEP-II and KEKB. (Figure 4.5) The positron production rate should be appropriate. An optimized injection scheme needs to be worked out in a conceptual design.

Large synchrotron tune helps to separate the instability modes for the transverse mode coupling instability (TMCI). However, TMCI is more important for the larger rings and lower energies, such as the case for the top-up ring at injection.

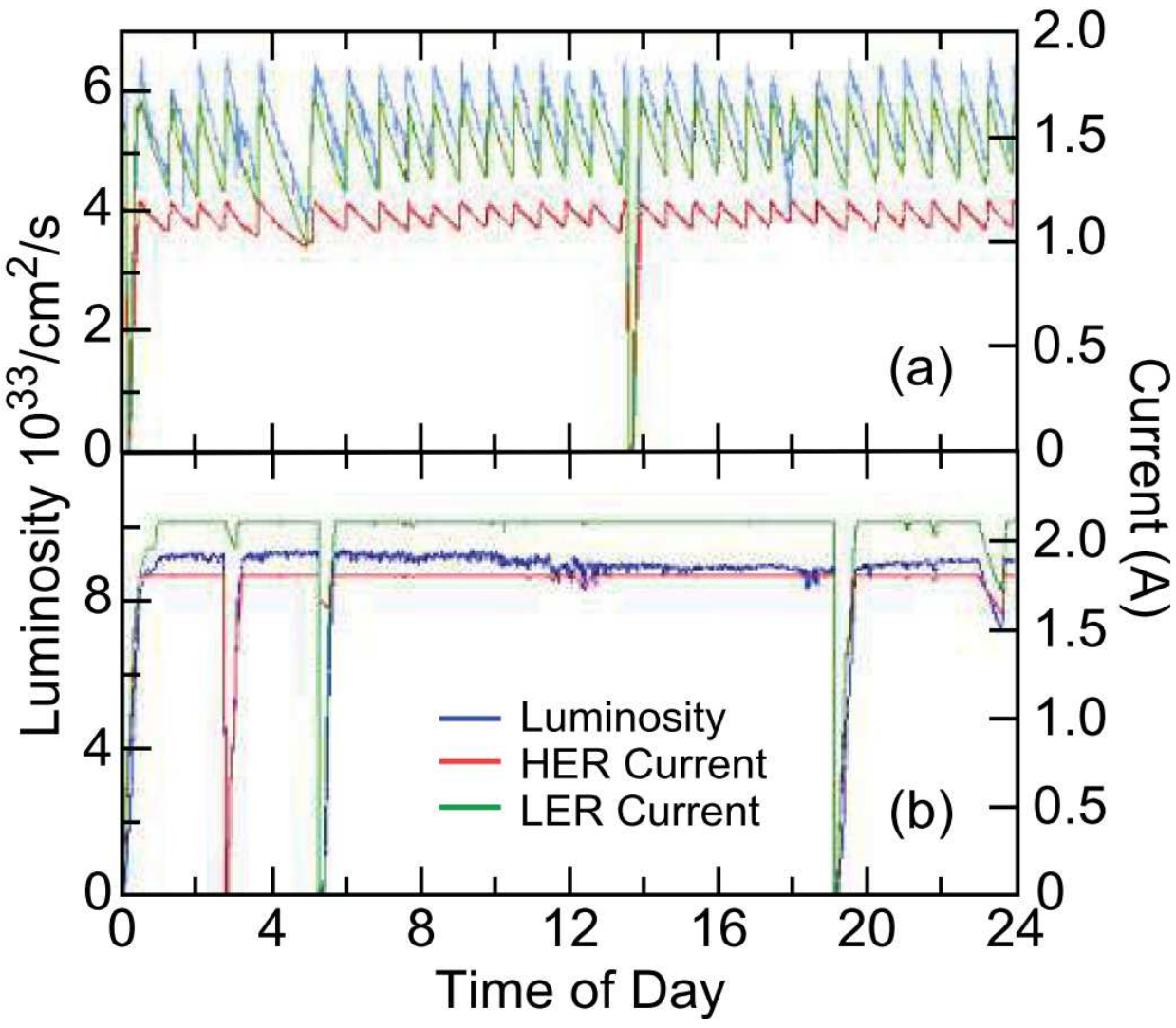

**Figure 4.5:** PEP-II operation: top – without top-up injection; bottom – with top-up injection.

#### 4.3.10 **Polarization**

In electron storage rings, polarization builds up by the Sokolov-Ternov effect. At high energies the beams will have difficulty reaching transverse polarizations due to their large energy spread. Therefore one should not expect to have polarized beams at 120 GeV or above in any of the proposed circular machines. Other means of beam energy calibration have to be devised. As seen in the physics section, this is not a big loss for the Higgs boson physics. Obtaining transverse polarization at the Z peak was achieved in LEP for the energy calibration, and this should not be difficult in the machines under discussion, especially since great efforts will be dedicated to achieving small vertical emittances. It would be of great benefit for the physics to have i) longitudinal polarization for colliding beams at the Z peak, and ii) availability of transverse polarization at the W threshold ($E_{beam}$ = 80-85 GeV) for calibration purposes. Both are uncertain at this point and need to be studied. Additional hardware such as movable spin rotators (as in the HERA) or polarization wigglers as in the LEP would probably be necessary.

# 5 Muon Collider

## 5.1 Introduction

A unique feature of a muon collider is its large cross-section in s-channel resonance for Higgs production. This cross-section is proportional to the mass squared of the colliding particles. Since the mass of a muon is about 200 times that of an electron, the cross-section of $\mu^+\mu^- \to H$ is about 40,000 times larger than that of $e^+e^- \to H$. This makes a muon collider particularly attractive to serve as a Higgs factory.

The muon collider has a number of other advantages as well as challenges when compared to other types of Higgs factory. They are summarized below:

- Advantages:
  - Large cross section σ ($\mu^+\mu^- \to H$) = 41 pb in s-channel resonance will compensate for low luminosity (to compare to $e^+e^- \to ZH$ at 0.2 pb)
  - Small physical size footprint
  - No synchrotron radiation problem
  - No beamstrahlung problem
  - Unique way for direct measurement of the Higgs line shape and total decay width Γ
  - Exquisite energy calibration
  - A path to very high energy lepton-lepton collisions
- Challenges:
  - Muon 4D and 6D cooling needs to be demonstrated
  - Need small center-of-mass energy spread (0.003%)
  - RF in a strong magnetic field
  - Background from constant muon decay
  - Significant R&D required towards end-to-end design and firming up luminosity figures

- Cost unknown (not much cheaper than a TeV muon collider)

Presently there are two main muon collider R&D programs in the world. One is the Muon Accelerator Program (MAP) in the U.S. coordinated by Fermilab. Another is the Muon Ionization Cooling Experiment (MICE) in the U.K. carried out by an international collaboration.

## 5.2 The Muon Collider as a Higgs Factory

The layout of a muon collider is shown in Figure 5.1. It consists of six stages:

- Proton driver – a high beam power (~4 MW), short beam pulse (~3 ns) proton accelerator. It can be either a rapid cycling synchrotron or a combination of a linac, an accumulator and a compressor.
- Target system – a mercury jet target and a solenoid for generating and capturing high flux pion beams.
- Front end – for pion decay to muon and phase rotating the muon beam for reducing the momentum spread.
- Cooling – a key part of the muon collider. It uses a complex magnet-RF system for ionization cooling to reduce the longitudinal and transverse emittance of the muon beam by several orders of magnitude.
- Acceleration – a number of options including linac, recirculating linac, rapid cycling synchrotron and FFAG.
- Collider ring – two muon beams, one $\mu^+$ another $\mu^-$, each of 63 GeV, colliding for Higgs production via the s-channel resonance.

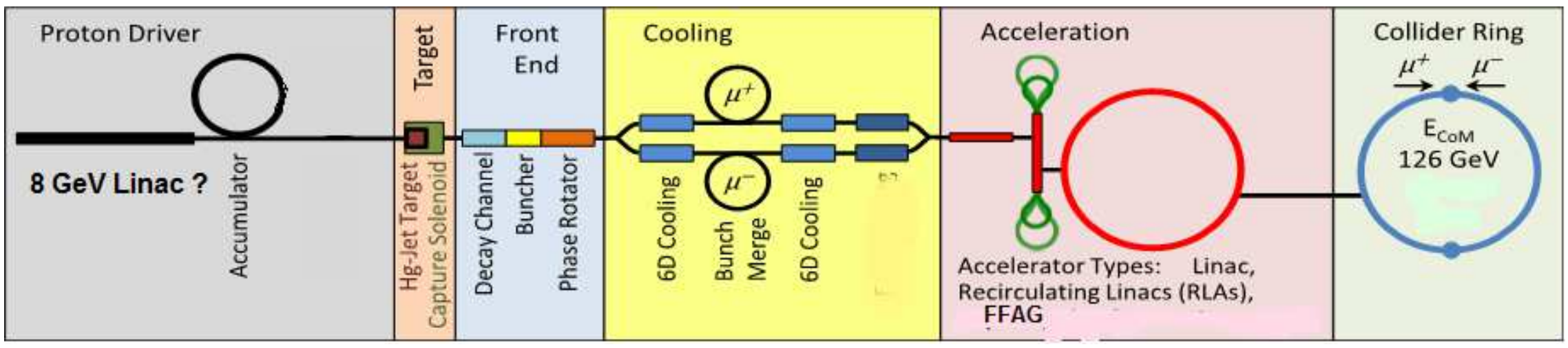

**Figure 5.1:** Layout of a muon collider.

The most demanding part is the cooling. As shown in Figure 5.2, the 6D cooling requires a total reduction of a factor of $10^6$ in beam emittance ($10^4$ in the transverse phase space and $10^2$ in the longitudinal phase space). And this has to be done rapidly (~msec) before the muons decay. The R&D involves a number of frontier technologies such as high temperature superconducting (HTS) high field magnets and high gradient low frequency superconducting RF (SRF). (It should be noted that for a Higgs factory, the requirement for the high field magnets is less demanding than that for a TeV-scale collider because the final cooling stage is not needed.) The breakdown of an RF cavity in a strong magnetic field is an example of the many challenging issues that a muon collider would have to solve before it can become a viable option for a Higgs factory. The Muon Accelerator Program (MAP) is making good progress in this direction.

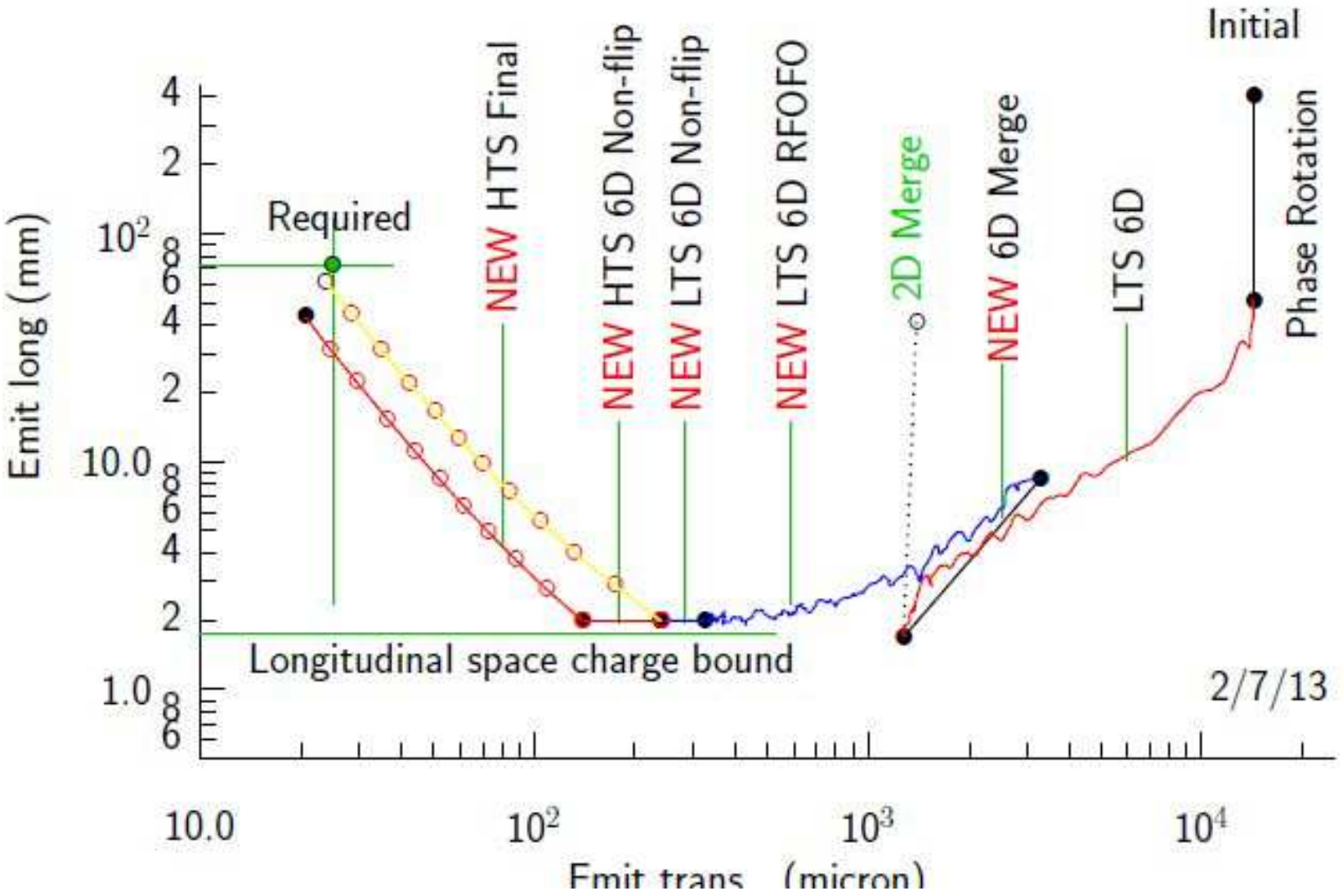

**Figure 5.2:** Muon cooling procedure.

When the muon collider serves as a Higgs factory, the longitudinal emittance is more important than the transverse one as high energy resolution of the muon beam is required for direct measurement of the Higgs total decay width Γ (see Sec. 2.4). Therefore, the cooling can end at the point where there is the smallest longitudinal emittance. The price to pay for this simplification is a lower luminosity due to larger transverse emittance. The design luminosity for a muon Higgs factory is $10^{31}$–$10^{32}$ $cm^{-2}s^{-1}$, which is about 2-3 orders of magnitude less than for an $e^+e^-$ collider. Fortunately, however, this is compensated by the large cross-section (~41 pb, to be compared to 0.24 pb for $e^+e^- \rightarrow ZH$) so the number of Higgs produced per year is comparable for the two types of Higgs factory.

As a muon is ~200 times heavier than an electron, the synchrotron radiation and beamstrahlung of a 63 GeV muon beam become irrelevant, and the ring size can be quite small (300 m), which is a big advantage of a muon collider.

An interesting feature of a muon beam is its energy can be calibrated precisely. With a polarization as small as 10%, the muon energy can be determined by measuring the oscillation frequency of the electrons (from muon decay) to an accuracy of ~0.1 MeV.

The parameters of a muon Higgs factory are listed in Table 8.3.

# 6 Photon Colliders

## 6.1 Introduction

The idea of a photon collider (γγ collider) dates back to 1981 and much significant work has been done since then. The photon collider is based on Inverse Compton Scattering ICS) as illustrated in Figure 6.1. The discovery of the “Higgs-like” boson brought about renewed interest to this concept. The advantage is that the cross section for $\gamma\gamma \rightarrow H$ is large and comparable to $e^+e^- \rightarrow ZH$ (~200 fb) but the required energy is much lower (63 GeV for a photon beam, corresponding to 80 GeV for an electron beam, compared to 120 GeV per electron beam in an $e^+e^-$ collider). This makes a photon collider an attractive option for either a low energy linear collider (80 GeV per electron beam) or a low energy circular collider (80 GeV per beam). Furthermore, for a photon collider there is no need for positrons and only one damping ring is needed.

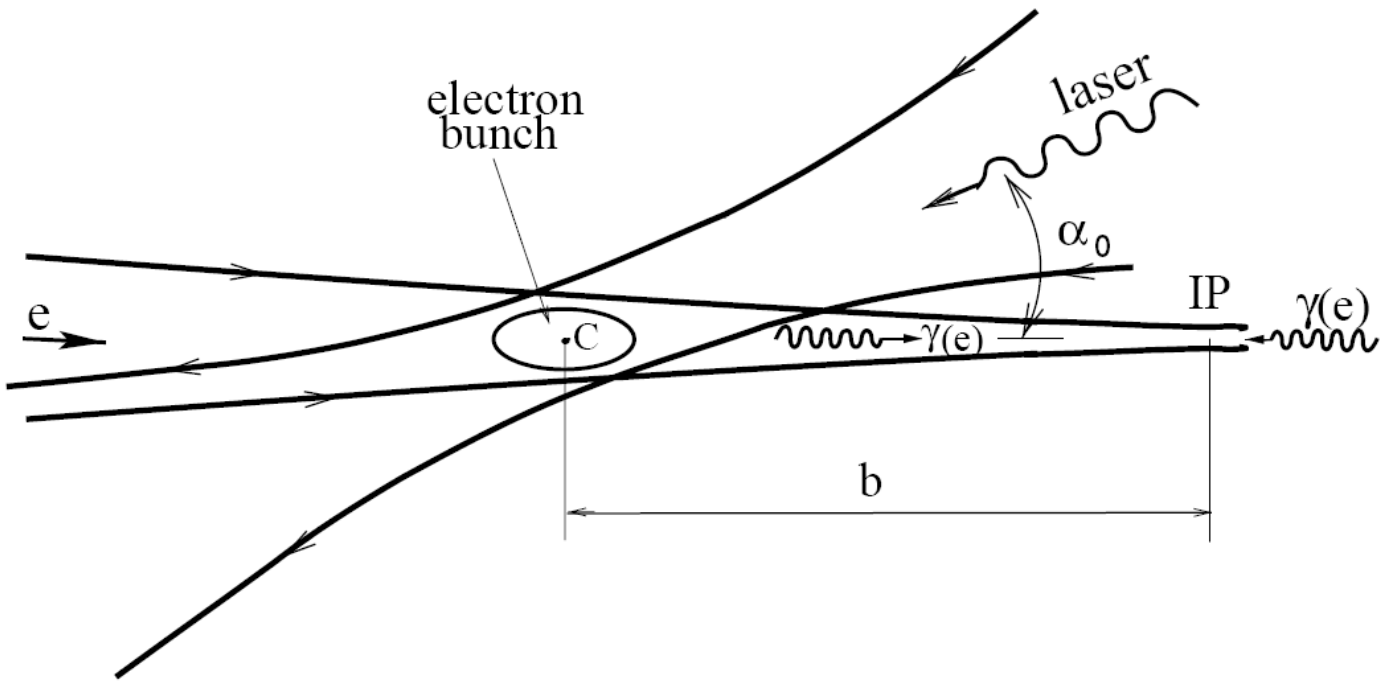

**Figure 6.1:** Illustration of the Inverse Compton Scattering.

Several possibilities of γγ colliders were presented in the workshop, including:
(1) ILC-based
(2) CLIC-based (CLICHÉ)
(3) NLC-type
(4) SLC-type
(5) SAPPHiRE
(6) Energy-Recovery Linac-based (This was not presented at the workshop, see Ref. [25])

From a physics point of view these concepts can be classified into two categories:
(a) Those which focus on s-channel Higgs production at $E\gamma\gamma_{CM}$ = 126 GeV;
(b) Those which are extendable to higher energies.

Among the above concepts (1), (2) and (3) belong to (b); (4), (5) and (6) to (a). The concepts in (a) include an arc for electrons of several tens of GeV so that they cannot be extended to higher energies in a realistic sense.

The concepts above can also be classified into two from the viewpoint of the distance between the bunches, namely,
(A) Those which are based on a normal-conducting linac so that the distance between bunches is of the order of a nanosecond;
(B) Those which are based on a superconducting linac with the bunch distance of order of a micro-second.

Among the above concepts (2) and (3) belong to (A), and (1), (4), (5), (6) to (B). ((4) belongs to (A) if a normal-conducting linac is used.) The requirements on the laser are quite different between these two categories.

The common features in all cases are:

- The primary electron beam must be highly polarized (≥ 80%) for obtaining a narrow energy spread of the photon beam.
- The laser flush energy at the conversion point must be several Joules (except (6), which requires one order of magnitude less), although the exact value depends on the electron bunch length.

Here we briefly describe these concepts. The proposed parameters are summarized in Table 8.4.

The linear collider-based ones – (1), (2) and (3) – are relatively obvious, so they are omitted here.

**SLC-type**: Uses 85 GeV pulsed normal-conducting or superconducting linac (or 45 GeV pulsed superconducting linac, twice-recirculating) to produce two ~80 GeV electron beams and the arcs of 1km radius to bend them as in SLC. No damping rings. For the laser, the ILC-type optical cavity or FEL is suggested.

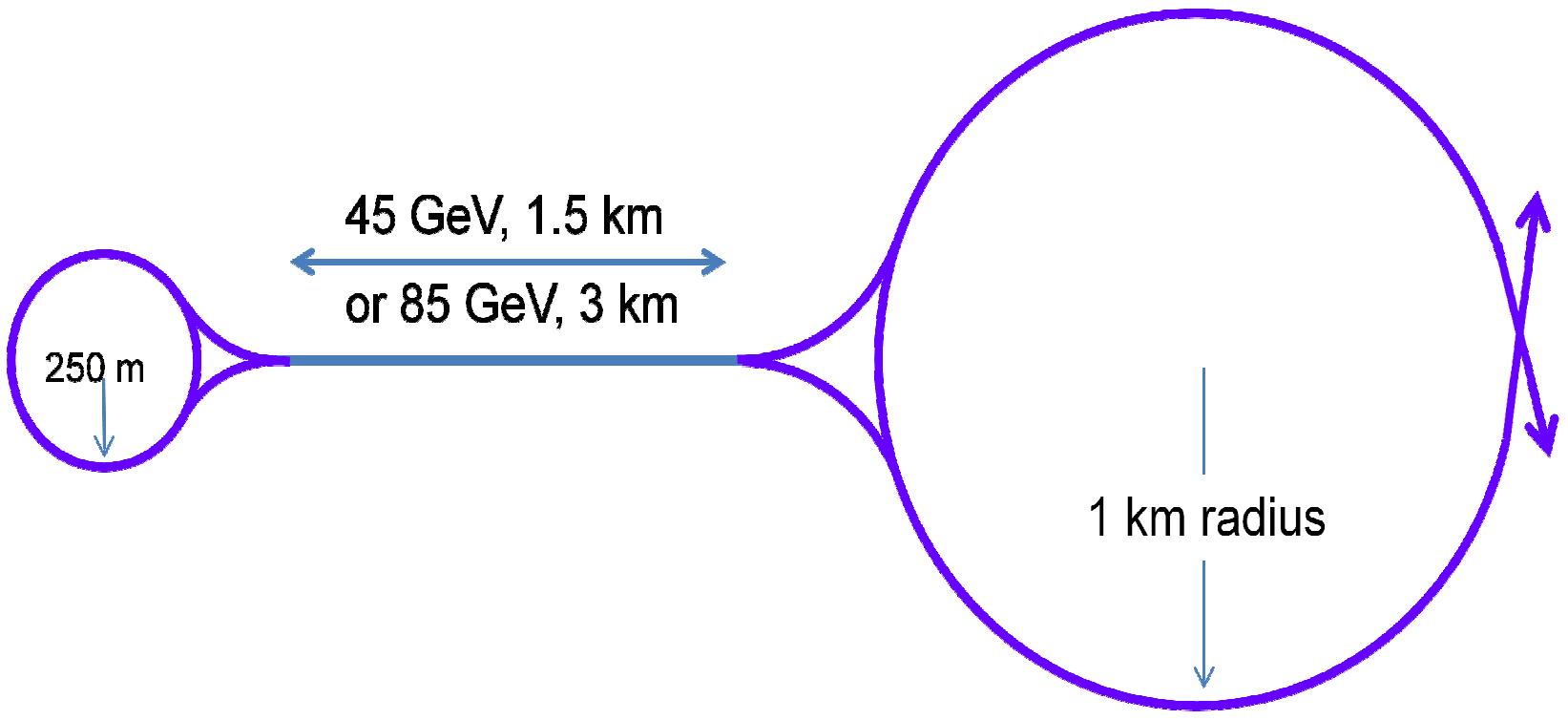

**Figure 6.2:** SLC-type photon collider.

**SAPPHiRE**: Uses recirculating CW linac (two 11 GeV linacs, 4 turns) with 80 GeV arcs and FNAL-A0-type flat-beam gun (but must be of higher intensity and polarized). No damping rings. The acceleration part of this scheme is an Energy Recovery Linac but the energy is not recovered when used as a photon collider.

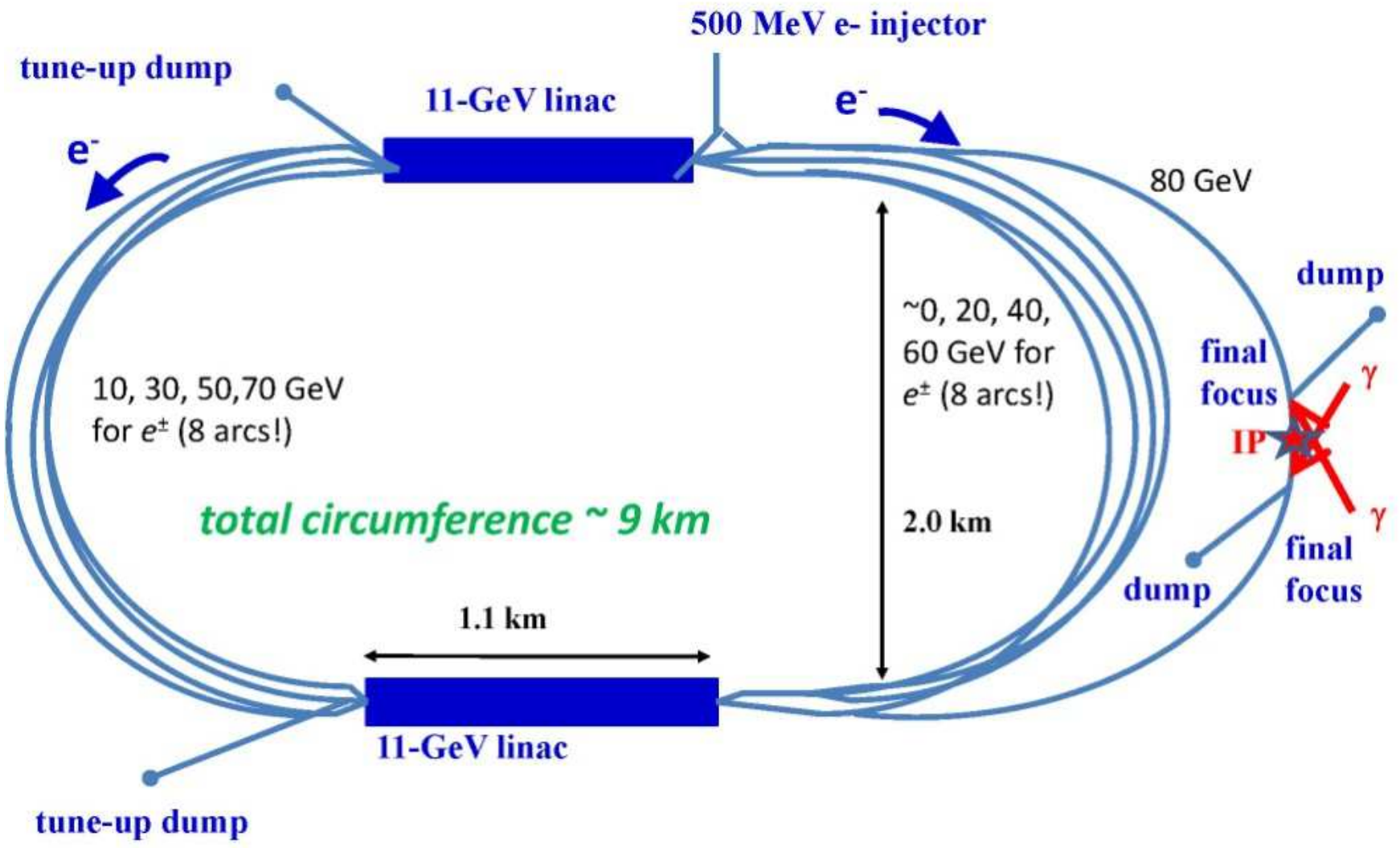

**Figure 6.3:** Layout of SAPPHiRE.

**Energy Recovery Linac-based**: Uses two 50 GeV CW SC linacs with 50 GeV arcs. The laser flush energy, and therefore the e-$\gamma$ conversion efficiency is much lower than in (5). The luminosity is partly restored by the much higher beam current. The merit of the low conversion efficiency is that the energy of the electrons that did not interact with laser photons can be recovered in the ERL configuration.

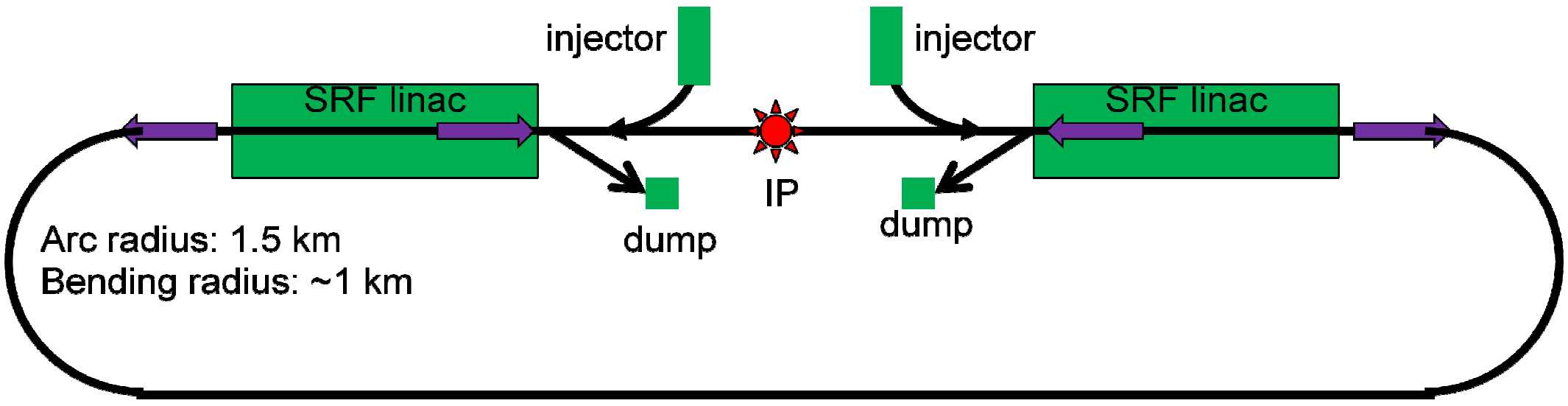

**Figure 6.4:** Energy-recovery linac-based photon collider.

Generally speaking, the $\gamma\gamma$ Higgs factory has the following advantages:

- The colliding $\gamma$ beams have high polarization.
- The required primary beam energy is relatively low (category (a)) so that the cost may be lower.
- It can be added to a linear collider (category (b)).
- The positron beam is not needed (category (a)).
- The damping ring, which is necessary for positron, may be omitted. Then, however, the electron gun must provide low emittance and high polarization.
- It allows access to CP property of the Higgs

The challenges are:

- The physics is not as comprehensive as it would be in a 240 GeV $e^+e^-$ collider. (for category (a))
- Background coming from the wide energy spectrum of the photons.
- The design of the interaction region is complex.
- High power laser technology is required.

## 6.2 Required R&D for Photon Colliders

Category (B) needs a laser repetition rate on the order of MHz with flush energy several Joules. This is unrealistic. Hence, an optical cavity to accumulate weak laser pulses is indispensable. The required parameter range are: Q-value O(1000), the optical path length O(100m), stored energy O(10J). The Q-value has already been reached but the present state-of-the art for the other two parameters is path length O(1m) and stored energy O(1mJ). Thus, intensive R&D is needed for the optical cavity. (Figure 6.5)

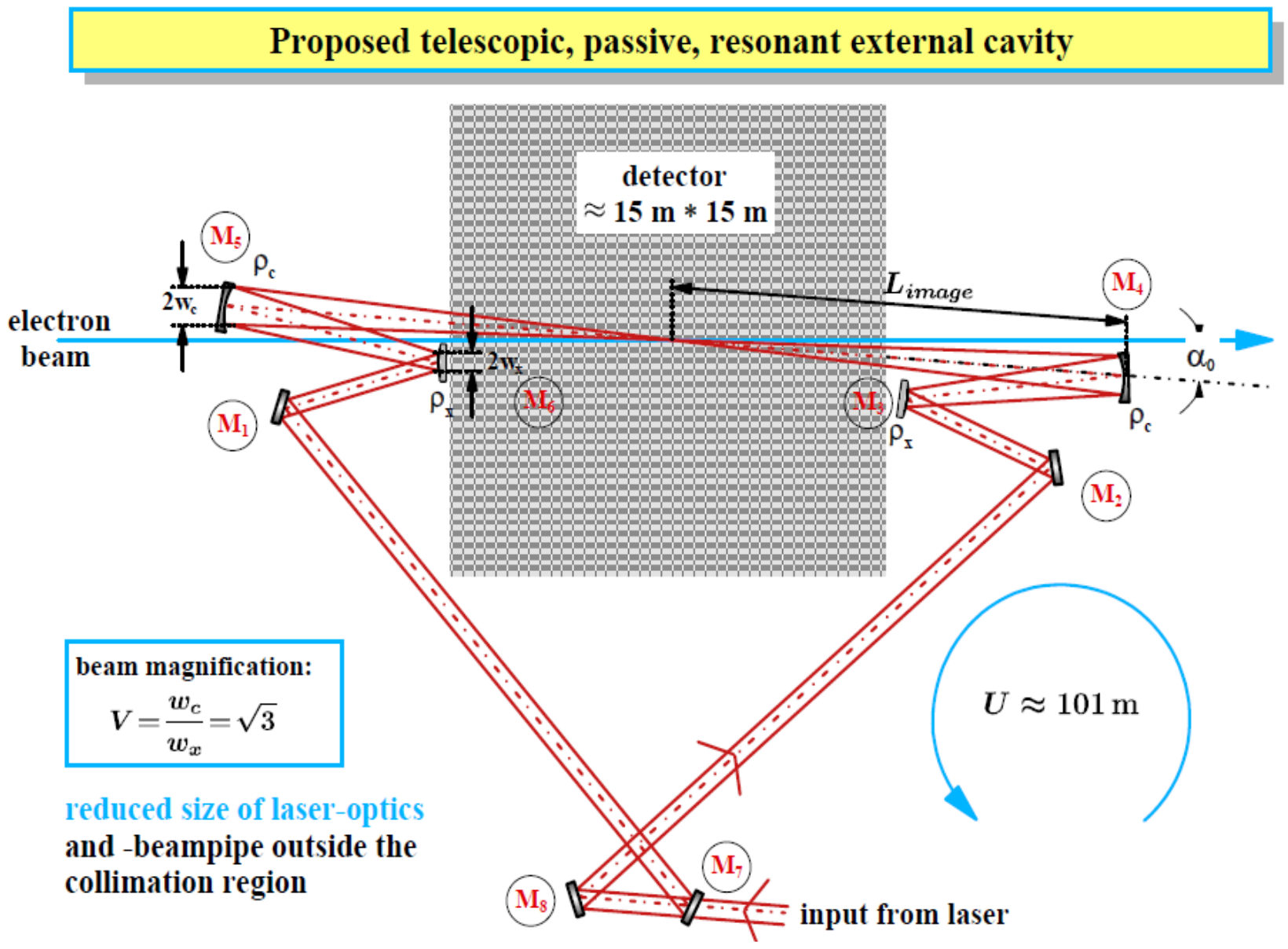

**Figure 6.5:** Sketch of an optical cavity in the interaction region.

Category (A) requires the laser flush energy O(kJ), split into some hundred sub-pulses of several Joules each, repeated at several tens of Hz. This sort of lasers may be available from inertia fusion technology such as LIFE. (Figure 6.6) However, it is not clear when they become available and when the technology to split the kJ pulse into sub-pulses becomes ready. For the R&D of laser systems required to implement a photon collider, the commitment of both the accelerator and laser communities is necessary.

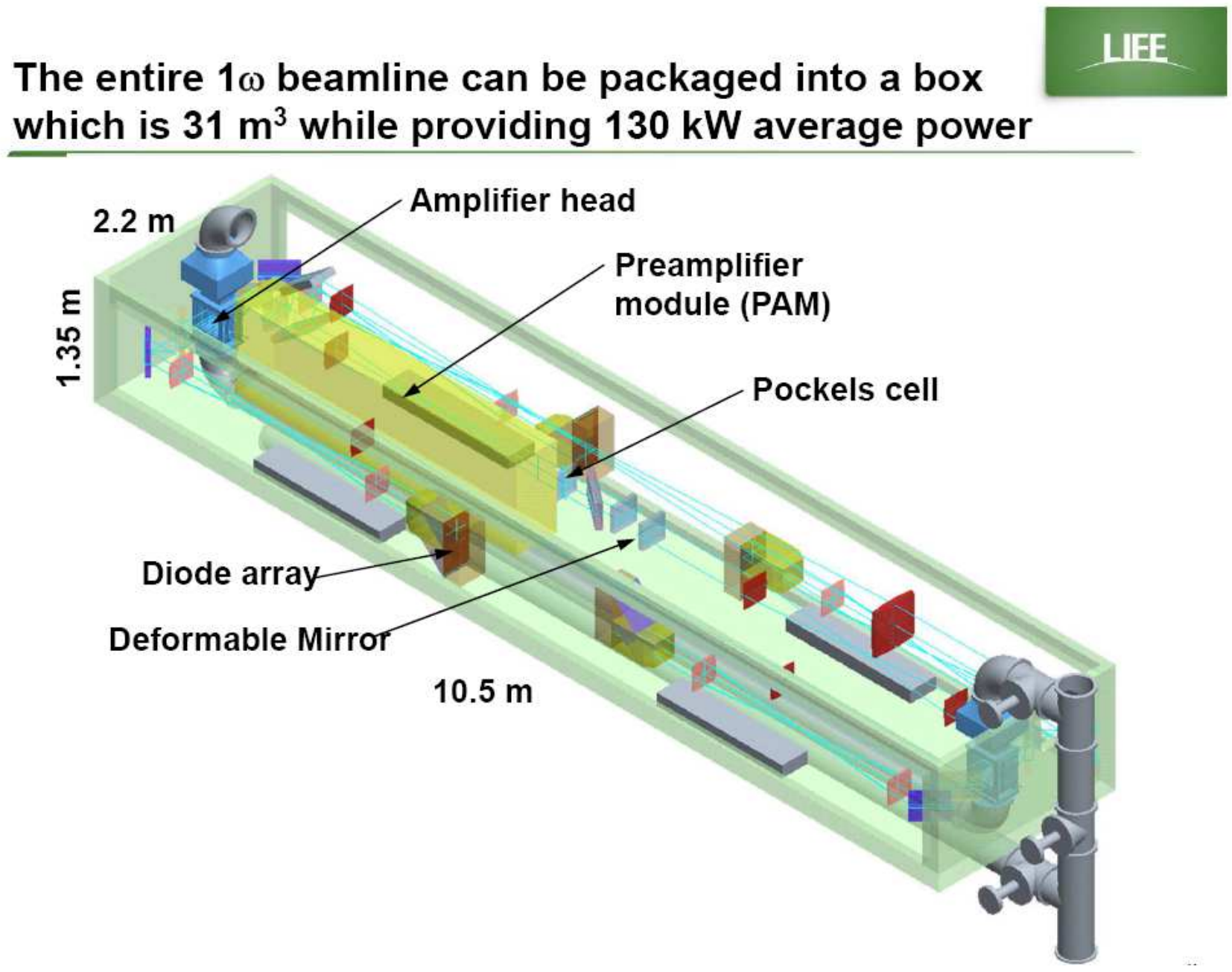

**Figure 6.6:** One of the 384 laser beam lines for the inertia fusion project LIFE at the Lawrence Livermore Laboratory.

Some of the proposals, including both (A) and (B), suggest the possibility of using an FEL. However, to produce several Joule pulse demands the energy extraction from electron beams with extremely high efficiency (≥ 10%). A tapered FEL is a possibility but this also requires R&D of many years including a demonstration. SAPPHiRE suggests a possibility of using the primary electron beams as the FEL driver, but this requires even more difficult R&D.

All the proposals of category (a) omit damping rings so they demand a low-emittance, highly polarized electron beam. The polarized RF gun is a possibility, but the achievement of this technology will require years of R&D. (4) allows larger emittance compared to (5) and (6) but requires larger bunch charge.

In all cases detailed design reports are still missing. The report should include:

- Design of the laser system
- Design of the polarized gun system
- More detailed design of the linac system for (4), (5) and (6). (In particular the cryogenics system for (6) may be very demanding.)
- More detailed design of the arcs and the entire geometry, which are to be reflected in the cost estimate.
- Cavity specification with power couplers
- Total power consumption
- Detailed design of the interaction region including the final-focus lattice and the path of the laser beams. (It is quite likely that the required laser flush energy would increase after detailed studies in particular for (4), (5) and (6). Relatively serious studies have already been done for (1), (2) and (3), though not at the level of a technical design.) (6) must also include the recovery path of the disrupted electrons.
- Background studies including realistic photon spectrum and the effects of the spent electrons.
- Estimated length of the R&D time. Those of category (a) are limited to s-channel Higgs production only. Therefore timely construction is essential. For those of category (b) there is a possibility of starting the machine as an $e^+e^-$ collider (around 240 GeV) and converting it later to a $\gamma\gamma$ collider when the technology becomes mature. (If one wants to start (b) as an s-channel Higgs machine for saving the initial cost, the situation is similar to (a) though polarized RF gun would not be needed.)

# 7 Acknowledgments

This workshop was endorsed by the ICFA and received strong support from the Fermilab Director's Office. The authors want to express their sincere gratitude to all the speakers and participants for their well-prepared excellent presentations and dynamic and stimulating discussions. The comments, suggestions and feedback we received from the world HEP community before, during and after the workshop were encouraging and most valuable for writing this report.

# 8 Appendices

## 8.1 Appendix 1 – Agenda

| Wednesday, November 14, Room One West | | |
|---|---|---|
| 08:00 – 09:00 | Registration | |
| | **Session 1: Introduction and physics** (Chair: Alain Blondel) | |
| 09:00 – 09:05<br>09:05 – 09:30<br>09:30 – 10:10<br>10:10 – 10:55 | Welcome (5')<br>Strategy for Higgs study (20')<br>Higgs at the LHC (30')<br>Higgs beyond the LHC – theories (30') | Pier Oddone (Fermilab)<br>Young-Kee Kim (Fermilab)<br>Fabio Cerutti (LBNL)<br>Chris Quigg (Fermilab) |
| 10:55 – 11:15 | Coffee (20') | |
| 11:15 – 12:00<br>12:00 – 12:30 | Higgs beyond the LHC – experiments (30')<br>Accelerators for a Higgs factory (20') | Patrick Janot (CERN)<br>Stuart Henderson (Fermilab) |
| 12:30 – 14:00 | Lunch | |
| | **Session 2: Linear e+e- Higgs factories** (Chair: Jie Gao) | |
| 14:00 – 14:45<br>14:45 – 15:30 | ILC as a Higgs factory (30')<br>CLIC as a Higgs factory (30') | Nick Walker (DESY)<br>Daniel Schulte (CERN) |
| 15:30 – 16:00 | Group photo & Coffee (30') | |
| 16:00 – 16:45<br>16:45 – 17:30 | SLC- & NLC-type Higgs factory (30')<br>Machine-detector interface for the ILC and CLIC (30') | Tor Raubenheimer (SLAC)<br>Marco Oriunno (SLAC) |
| 17:30 – 19:30 | Reception (Wilson Hall, 2nd floor) | |

| Thursday, November 15, Room One West | | |
|---|---|---|
| | **Session 3: Circular e+e- Higgs factories** (Chair: Daniel Schulte) | |
| 09:00 – 09:40 | LEP3 and TLEP (25’) | Frank Zimmermann (CERN) |
| 09:40 – 10:05 | SuperTristan (15’) | Katsunobu Oide (KEK) |
| 10:05 – 10:30 | Fermilab site-filler (15’) | Tanaji Sen (Fermilab) |
| 10:30 – 11:00 | Coffee (30’) | |
| 11:00 – 11:25 | IHEP Higgs factory (15’) | Qing Qin (IHEP) |
| 11:25 – 12:05 | LBNL/SLAC ring and lattice issues (25’) | Yunhai Cai (SLAC) |
| 12:05 – 12:30 | Topping up injection (15’) | John Seeman (SLAC) |
| 12:30 – 14:00 | Lunch | |
| | **Session 4: Limits for circular e+e- colliders** (Chair: Alex Chao) | |
| 14:00 – 14:30 | Beamstrahlung – calculations and cure (20’) | Valery Telnov (BINP) |
| 14:30 – 15:00 | Beamstrahlung – simulations (20’) | Marco Zanetti (MIT) |
| 15:00 – 15:30 | Scaling law (20’) | Kaoru Yokoya (KEK) |
| 15:30 – 16:00 | Coffee (30’) | |
| 16:00 – 16:30 | Beam-beam tune shift (20’) | Jie Gao (IHEP) |
| 16:30 – 17:00 | Synchrotron radiation – RF (20’) | Andy Butterworth (CERN) |
| 17:00 – 17:30 | Synchrotron radiation – vacuum (20’) | Nadine Kurita (SLAC) |
| 18:30 – 20:30 | Dinner (Users Center – Chez Leon) | |

| Friday, November 16, Room One West | | |
|---|---|---|
| | **Session 5** (Chair: Weiren Chou) | |
| | **5A: Low emittance rings** | |
| 09:00 – 09:45 | Light sources (30’) | Riccardo Bartolini (Diamond) |
| 09:45 – 10:30 | Colliders (30’) | Yoshihiro Funakoshi (KEK) |
| 10:30 – 11:00 | Coffee (30’) | |
| | **5B: Muon collider as a Higgs factory** | |
| 11:00 – 11:20 | Physics of $\mu\mu \rightarrow$ Higgs (15’) | Tao Han (U. of Pittsburgh) |
| 11:20 – 12:05 | Muon collider (30’) | David Neuffer (Fermilab) |
| 12:05 – 12:30 | Background and machine-detector interface (15’) | Ron Lipton (Fermilab) |
| 12:30 – 13:45 | Lunch | |
| | **Session 6** | |
| | **6A: $\gamma\gamma$ collider as a Higgs factory** (Chair: Kaoru Yokoya) | |
| 13:45 – 14:10 | Physics of $\gamma\gamma \rightarrow$ Higgs (15’) | Mayda Velasco (Northwestern U.) |
| 14:10 – 14:45 | $\gamma\gamma$ collider (25’) | Tohru Takahashi (Hiroshima U.) |
| 14:45 – 15:10 | Laser for CLIC-based $\gamma\gamma$ collider (15’) | Andy Bayramian (LLNL) |
| 15:10 – 15:30 | SAPPHIRE (15’) | Frank Zimmermann (CERN) |
| 15:30 – 16:00 | Wine & Cheese (30’) | |
| | **6B: Summary Talks (Joint with the Wine & Cheese Seminar)** (Chair: John Campbell) | |
| 16:00 – 16:25 | Higgs factory – physics (25’) | Alain Blondel (U. of Geneva) |
| 16:25 – 17:00 | Higgs factory – accelerators (35’) | Weiren Chou (Fermilab) |
| 17:00 | Adjourn | |

## 8.2 Appendix 2 – Parameter Comparison Tables

At the workshop a set of parameter tables were compiled with input from the proponents of each proposal. There are four parameter tables, one for each category of Higgs factory, namely,

- Linear $e^+e^-$ colliders
- Circular $e^+e^-$ colliders
- Muon collider
- Photon colliders

Each table contains two parts:

- Top level parameters, including:
    - Center-of mass energy
    - Luminosity
    - Number of interaction points (IP)
    - Number of Higgs per year per IP
    - Machine size (length or circumference)
    - Power consumption
    - Polarization
    - Energy upgrade limit

  These parameters are common to all Higgs factories and can be used for cross comparison.
- Other important parameters:
  These can be different for different categories of Higgs factory. They can be used for comparing different machines in the same category.

These tables were provided by the workshop presenters except for some obvious corrections and items derived from the data provided.

**Table 8.1:** Parameters of Linear $e^+e^-$ Colliders (Note: The CLIC 250 GeV numbers are for the 500 GeV machine operating at 250 GeV)

| | | Linear e+e- collider | | | | | | |
|---|---|---|---|---|---|---|---|---|
| | | **ILC** | | | **CLIC** | | | **X-band** |
| | | | | | **A+** | **A** | **A/B** | **Klystron-based** |
| **Top Level Parameters** | | | | | | | | |
| Energy (center of mass) | GeV | 250 | 500 | 1000 | 250 | 500 | 3000 | |
| Luminosity (per IP) | $10^{34}$ $cm^{-2}$ $s^{-1}$ | 0.75 | 1.8 | 4.9 | 1.37 | 2.3 | 5.9 | |
| No. of IP | | 1 | 1 | 1 | 1 | 1 | 1 | |
| No. of Higgs per year (per IP) | 1000 | 23 | 49 | | 34 | 44 | 446 | |
| Size (length or circumference) | km | 21 | 31 | 48 | 13.2 | 13.2 | 48.3 | |
| P(wall) | MW | 128 | 162 | 301 | 235 | 272 | 589 | |
| Polarization | | | | | | | | |
| e- | % | 80 | 80 | 80 | 80 | 80 | 80 | |
| e+ | % | 30 | 30 | 30 | 0 | 0 | 0 | |
| | | | | | | | | |
| | | | | | | | | |
| **Other Important Parameters** | | | | | | | | |
| Geometric luminosity | $10^{34}$ $cm^{-2}$ $s^{-1}$ | 0.37 | 0.75 | 2.61 | 0.82 | 1.42 | 4.29 | |
| Pinch enhanced luminosity | $10^{34}$ $cm^{-2}$ $s^{-1}$ | 0.75 | 1.8 | 4.9 | 1.37 | 2.3 | 5.9 | |
| Linac accelerating gradient | MV/m | 31.5 | 31.5 | 31.5/45 | 40 | 80 | 100 | |
| No. of particles per bunch | 10^10 | 2 | 2 | 1.74 | 0.34 | 0.68 | 0.372 | |
| Number of bunches per pulse | | 1312 | 1312 | 2450 | 842 | 354 | 312 | |
| Bunch distance | ns | 554 | 554 | 366 | 0.5 | 0.5 | 0.5 | |
| Average current / beam | mA | 0.0210 | 0.0210 | 0.0273 | 0.0229 | 0.0193 | 0.0093 | |
| Peak current / beam | A | 0.0058 | 0.0058 | 0.0076 | 1.0894 | 2.1787 | 1.1919 | |
| Pulse repetition rate | Hz | 5 | 5 | 4 | 50 | 50 | 50 | |
| Beam power per beam | MW | 2.63 | 5.25 | 13.66 | 2.87 | 4.82 | 13.95 | |
| Normalized emittance $\varepsilon_{x,n}$ | mm-mrad | 10 | 10 | 10 | 0.66 | 2.4 | 0.66 | |
| Normalized emittance $\varepsilon_{y,n}$ | mm-mrad | 0.035 | 0.035 | 0.03 | 0.025 | 0.025 | 0.02 | |
| $\beta_x$ IP | mm | 13 | 11 | 11 | 8 | 8 | 4 | |
| $\beta_y$ IP | mm | 0.41 | 0.48 | 0.23 | 0.1 | 0.1 | 0.07 | |
| $\sigma_x$, IP | nm | 729 | 474 | 335 | 150 | 200 | 40 | |
| $\sigma_y$, IP | nm | 7.66 | 5.90 | 2.70 | 3.2 | 2.3 | 1 | |
| $\sigma_z$, IP | mm | 0.3 | 0.3 | 0.225 | 0.072 | 0.072 | 0.044 | |
| sigma_E IP (electron) | % | 0.19 | 0.124 | 0.085 | 0.3 | 0.3 | 0.3 | |
| sigma_E IP (positron) | % | 0.152 | 0.07 | 0.047 | 0.3 | 0.3 | 0.3 | |
| Full crossing angle | mrad | 14 | 14 | 14 | 18.6 | 18.6 | 20 | |
| Average number of photons | | 1.176 | 1.700 | 2.250 | 0.7 | 1.3 | 2.1 | |
| $\delta_B$ beam-beam | % | 0.953 | 4.500 | 10.500 | 1.5 | 7 | 28 | |
| Upsilon (average) | | 0.0201 | 0.0616 | 0.203 | 0.0515 | 0.207 | 5.49 | |

**Table 8.2:** Parameters of Circular $e^+e^-$ Colliders

| | | Circular e+e- collider | | | | | | | |
|---|---|---|---|---|---|---|---|---|---|
| | | **LEP3** | **TLEP** | | **Super-TRISTAN** | **Fermilab Site-filler** | **IHEP Ring** | | **SLAC/LBNL Ring** |
| | | | | | | | **IHEP-50km** | **IHEP-80km** | |
| **Top Level Parameters** | | | | | | | | | |
| Energy (center of mass) | GeV | 240 | 240 | 350 | 240 | 240 | 240 | 240 | 240 |
| Luminosity (per IP) | $10^{34}$ $cm^{-2}$ $s^{-1}$ | 1 | 4.9 | 0.65 | 1 | 0.52 | 2.5 | 3.85 | 1 |
| No. of IP | | 2 (4 ) | 2 | 4 | 1 | 1 | 2 | 1 | 1 |
| No. of Higgs per year (per IP) | 1000 | 20 | | 100 | | 13 | 100 | 200 | |
| Size (length or circumference) | km | 26.7 | 81 | 81 | 40 | 16 | 49.78 | 69.88 | 26.7 |
| P(wall) | MW | 200 | 200 | 200 | 100 | 200 | 300 | 300 | 200 |
| Polarization | | | | | | | | | |
| e- | | 0 | 0 | 0 | 0 | 0 | 0 | 0 | 0 |
| e+ | | 0 | 0 | 0 | 0 | 0 | 0 | 0 | 0 |
| | | | | | | | | | |
| **Other Important Parameters** | | | | | | | | | |
| Bending radius | km | 2.6 | 9 | 9 | 5.4 | 1.753 | 6.2 | 7.8 | 2.6 |
| Ne | $10^{10}$ per bunch | 100 | 50 | 75 | 67 | 80 | 70 | 60 | 8 |
| nb (number of bunches) per beam | | 4 | 80 | 12 | 8 | 2 | 26 | 52 | 50 |
| I(beam) | mA | 7.2 | 24.3 | 5.4 | 6.5 | 5 | 17.5 | 21.3 | 7.2 |
| ΔE(synch) | GeV/turn | 6.99 | 2.1 | 9.3 | 3.5 | 10.5 | 3 | 2.35 | 6.99 |
| P(synch) per beam | MW | 50 | 50 | 50 | 22.5 | 50 | 51.8 | 50 | 50 |
| Critical energy of synch. rad. | MeV | 1.47 | 0.43 | 1.32 | 0.71 | 2.19 | 0.62 | 0.49 | 1.47 |
| εx,n | mm-mrad | 5870 | 2210 | 6850 | 9400 | 5321 | 3053 | 3358 | 1010 |
| εy,n | mm-mrad | 23 | 12 | 34 | 9.4 | 27 | 21.14 | 16.67 | 5.05 |
| beta_x IP | mm | 200 | 200 | 200 | 200 | 200 | 280 | 200 | 50 |
| beta_y IP | mm | 1 | 1 | 1 | 1 | 2 | 1 | 1 | 1 |
| σx, IP | nm | 71000 | 43000 | 63000 | 89000 | 67319 | 60000 | 53479 | 14663 |
| σy, IP | nm | 320 | 220 | 320 | 63 | 476 | 300 | 266 | 2.64 |
| σz, IP | mm | 3.1 | 1.7 | 2.5 | 1.2 | 2.85 | 1.6 | 1 | 1.5 |
| sigma_E, IP | % | 0.23 | 0.15 | 0.22 | 0.14 | 0.28 | 0.13 | 0.12 | 0.2 |
| Full crossing angle | mrad | 0 | 0 | 0 | 0 | 0 | 0 | 0 | 0 |
| Beam lifetime due to radiative Bhabha | sec | 1080 | 1920 | 3240 | | 1080 | 960 | 600 | |
| b-b tune shift x | | 0.09 | 0.1 | 0.05 | 0.032 | 0.067 | 0.102 | 0.08 | 0.036 |
| b-b tune shift y | | 0.08 | 0.1 | 0.05 | 0.083 | 0.095 | 0.073 | 0.08 | 0.07 |
| Damping partition number (x) | | 1.5 | 1 | 1 | 1 | 1.5 | 1 | 1 | 1 |
| Damping partition number (y) | | 1 | 1 | 1 | 1 | 1 | 1 | 1 | 1 |
| Damping partition number (z) | | 1.5 | 2 | 2 | 2 | 1.5 | 2 | 2 | 2 |
| Longitudinal damping time | turns | 26 | 57 | 19 | 35 | 11 | 41 | 51 | 15 |
| RF Voltage | GV | 12 | 6 | 12 | 8.3 | 12 | 9 | 12 | 12 |
| Momentum compaction | | | | | | | | | 0.000024 |
| Synchrotron oscillation tune | | 0.19 | 0.12 | 0.12 | 0.23 | 0.192 | 0.392 | 0.364 | 0.135 |
| Average number of photons | | 0.6 | 0.5 | 0.51 | 3.2 | 0.36 | 0.5 | 0.48 | 0.24 |
| δB beam-beam | % | 0.03 | 0.035 | 0.035 | 0.02 | 0.022 | 0.038 | 0.057 | 0.0088 |
| Upsilon(max) | | 0.00231 | 0.00348 | 0.00353 | 0.00320 | 0.00212 | 0.00371 | 0.00570 | 0.00186 |

**Table 8.3:** Parameters of Muon Collider

| | | Muon collider | |
|---|---|---|---|
| | | **Muon-Low L** | **Muon-High L** |
| **Top Level Parameters** | | | |
| Energy (center of mass) | GeV | 126 | 126 |
| Luminosity (per IP) | $10^{34}$ $cm^{-2}$ $s^{-1}$ | 0.001 | 0.01 |
| No. of IP | | 2 | 2 |
| No. of Higgs per year (per IP) | 1000 | 5 | 50 |
| Size (length or circumference) | km | 0.3 | 0.3 |
| P(wall) | MW | 100 | 125 |
| Polarization | | | |
| $\mu+$, $\mu-$ | | 10% | 10-20% |
| Energy upgrade limit | | 10 TeV | 10 TeV |
| | | | |
| **Other Important Parameters** | | | |
| Bending radius | km | 0.04 | 0.04 |
| Nmu | $10^{10}$ per bunch | 200 | 500 |
| nb (number of bunches) | | 1 | 1 |
| Normalized emittance $\varepsilon$x,n | mm-mrad | 400 | 200 |
| Normalized emittance $\varepsilon$y,n | mm-mrad | 400 | 200 |
| $\beta$x IP | mm | 60 | 40 |
| $\beta$y IP | mm | 60 | 40 |
| $\sigma$x, IP | nm | 200000 | 120000 |
| $\sigma$y, IP | nm | 200000 | 120000 |
| $\sigma$z, IP | mm | 60 | 40 |
| sigma_E, IP | % | 0.003 | 0.003 |
| Luminosity life time | sec | 0.00066 | 0.00066 |
| Repetition rate | Hz | 15 | 15 |

## **Table 8.4:** Parameters of Photon Colliders

| | | **γγ collider** | | | | | | |
|---|---|---|---|---|---|---|---|---|
| | | **ILC-based** | | **CLIC-based** | | **Recirculating linac-based** | **SLC-based** | **Thin Target** |
| | | **ILC (x=4.46)** | **ILC (x=1.97)** | **CLIC (x=4.46)** | **CLIC (x=4.46)** | | | |
| **Top Level Parameters** | | | | | | | | |
| Energy (center of mass) | GeV | 126 (lum. peak) | 126 (lum.peak) | 126 | 126 | 132 (max γγ) | 130 | |
| Luminosity (per IP) | $10^{34}$ $cm^{-2}$ $s^{-1}$ | 0.03 | 0.12 | 0.36 | 0.65 | 0.06 | 1 | 0.15 |
| definition of luminosity | | γγ > 125 GeV | γγ > 125 GeV | γγ > 0.6ECM | γγ > 0.6ECM | γγ > 125 GeV | | γγ > 0.6ECM |
| No. of IP | | 1 | 1 | 1 | 1 | 1 | 1 | 1 |
| No. of Higgs per year (per IP) | 1000 | 5 | 10 | | | 10 to 20 | 5 | |
| Size (length or circumference) | km | ~14? | ~16? | | | 9 | 12 | |
| P(wall) | MW | ~100? | ~100? | 150? | 300? | 100 | 150 | |
| Polarization | | | | | | | | |
| e- | | 80% | 80% | 80% | 80% | 90% | 80% | 80% |
| γ | | 93% (lum. peak) | 86% (lum. peak) | 100% | 100% | 90% | | |
| Energy upgrade limit | | 0.8Ee- (typical) | 0.8Ee- (typical) | | | ? | 260 | |
| | | | | | | | | |
| **Other Important Parameters** | | | | | | | | |
| Drive electron energy | GeV | 83 | 110 | 80? | 80? | 80 | 80 | 100 |
| Ne | $10^{10}$ per bunch | 2 | 2 | 0.4 | 0.68 | 1 | 5 | 1.5 |
| nb (number of bunches per pulse) | | 2860 | 2860 | 1694 | 2124 | CW | 1000 | CW |
| bunch distance | μs | 0.33 | 0.33 | 0.5(154) x 4(11) | 0.5(354) x 24(6) | 5 | 1 | 1 |
| electron I(beam) | mA | 0.045 | 0.045 | 0.11 | | 0.3 | 10 | 2.4 |
| Pulse repetition rate | Hz | 5 | 5 | 100 | 50 | CW | 10 | CW |
| electron P(beam) | MW | 3.8 | 5.0 | 8.6 | | 50 | 8 | 480 |
| **Electron beam** | | | | | | | | |
| εx,n | mm-mrad | 10 | 10 | 1.4 | 1.4 | 5 | 6 | 5 |
| εy,n | mm-mrad | 0.03 | 0.03 | 0.05 | 0.05 | 0.5 | 5 | 0.5 |
| beta_x CP | mm | 4.5 | 4.56 | | | 5 | | |
| beta_y CP | mm | 5.3 | 6.0 | | | 100 | | |
| beta_x IP | | | | | | 5 | 0.5 | |
| beta_y IP | | | | | | 0.1 | 0.5 | |
| σx, CP | nm | 535 | 460 | | | 400 | 140 | 385 |
| σy, CP | nm | 32 | 29 | 0.3 | 0.3 | 440 | 125 | 320 |
| σz, CP | mm | 0.35 | 0.35 | | | 0.03 | 200 | 0.1 |
| sigma_E IP | % | 0.22 in ML | 0.22 in ML | 0.351 | 0.351 | <0.1 | 0.5 | |
| **Laser beam** | | | | | | | | |
| wavelength | μm | 0.351 | 1.054 | 0.351 | | 0.35 | | 0.395 |
| Flush energy | J | 9 | 9 | | | 5 | | 0.075 |
| Rayleigh length | mm | 0.63 | 0.63 | | | 0.3 | | |
| σx, CP | nm | 4200 | 7300 | | | 2800 | | |
| σy, CP | nm | 4200 | 7300 | | | 2800 | | |
| σz, CP | mm | 0.45 | 0.45 | | | 0.15 | | |
| IP<->CP distance | mm | 1.4 | 1.5 | | | 0.6 | | 2 |
| Laser-beam crossing angle | mrad | | | | | 0 | | 20 |
| **γ beam** | | | | | | | | |
| n_gamma | | 1.0e10 (primary) | 1.6e10 (primary) | 9.60E+10 | 1.63E+11 | 8.30E+09 | | 3.20E+09 |
| σx, IP | nm | 480 | 430 | | | 400 | | |
| σy, IP | nm | 10 | 7 | | | 18 | | |

## 8.3 Appendix 3 – Timelines

Listed below are our best estimates or, sometimes, "guesstimates" of the timeline of various proposals for a Higgs factory. For each proposal, we shall identify a date of completion of its experimental program (not a date for completion of the facility).

1. **LHC**

The LHC dates are the most readily available. The date for completion of "nominal LHC" at 14 TeV with ~300 $fb^{-1}$ is assumed to be 2021, which is the beginning of the LHC long shut down of 2022-2023. The High Luminosity running is assumed to last until 2030 for an integrated luminosity of 3000 $fb^{-1}$. There is a possibility that this may be extended by a few years. A further extension to even higher energies by replacing the LHC magnets by 16-20 T peak field magnets in situ (HE-LHC) would take quite a few years after that date; similarly the timescale to fill an 80 km long tunnel with 20 T magnets (SHE-LHC) is also very uncertain.

2. **Linear $e^+e^-$ colliders**

The ILC dates were hinted at in the statement of interest of Japanese colleagues at the European Strategy ESPP workshop in Krakow (September 2012). It is stated that the project is ready to go. The Technical Design Report with cost estimate will be published in June 2013. Assuming a decision is taken within the next couple of years, and after due process it is assumed in the TDR that the construction time is around 9 years, this having to include the construction of a new lab infrastructure; the main uncertainty here comes from the availability of the appropriate funding level. A commissioning time of one year is assumed. Thus, it is assumed the physics run could begin around 2025. Five years of continuous running at 250 GeV $E_{CM}$ is, according to the ILC documents, necessary to reach 250 $fb^{-1}$, by a date which we estimate to be around 2030. The upgrade and additional running at 350 GeV, 500 GeV and 1 TeV lead to a completion date around 2040-2045.

The current planning for the CLIC foresees a development phase until 2016, which allows taking a decision on the next high energy frontier project to be taken in 2016-2017, based on the LHC results. In the preparation phase until 2022 the technical designs would be finalized, the industrial procurement prepared and the site authorisation would be obtained to be ready for construction starting from 2023 and finishing by 2030. So data could be taken as the LHC programme reaches completion.

3. **Circular $e^+e^-$ colliders**

Compared with the linear colliders, the circular Higgs factory proposals are much less construction ready. The ILC, for example, has its TDR completed with accurate cost estimate. In contrast, the circular proposals have yet to form the study groups that line up to produce CDRs. Cost estimates for the circular options are even further in the future.

As the circular Higgs factories are in their infancy, it is difficult to make reasonable predictions. Consider the cases of LEP3 and TLEP. It took 13 years from the first CERN yellow report in 1976 to the LEP start-up, of which 6 years from ground-breaking to first collisions. The first step is to produce a conceptual design study report. Given the significant interest raised at the workshop and the available worldwide expertise this should take two years, reaching a decision point following which 2-3 more years would be necessary for a TDR. The following is a possible CERN scenario; we expect a similar time scale for other circular Higgs factory proposals.

Assuming a decision in the few years following the next CERN strategy update in 2018, construction could possibly start in 2021 when the main components of the LHC upgrade are acquired. Construction of an 80 km tunnel should be at least similar to that of the LEP ring (tunnelling technology has made progress in recent years), thus at least 6 years. The components of the machine need to be developed and built in parallel. In a new ring the installation and commissioning is assumed to be straightforward and takes two years. Thus data-taking could possibly start around 2030 leading to 5 years worth of precision results not earlier than 2035. A full program on TLEP including polarized beams at the Z peak and exploration at the WW and ttbar thresholds would probably take another 10-15 years. The increase of complexity (double ring, larger RF system, longer vacuum chamber) with respect to LEP is a source of significant uncertainty in this number. LEP3 in the LHC tunnel is a fall back in case the funding for the larger projects does not come forth, but because of the need to install the machine in the LHC tunnel itself, it is difficult to expect an earlier time scale than that of TLEP. A new ring of similar circumference as LEP3 could probably be built faster at another site.

### 4. Muon collider

For the muon collider, the MAP program will deliver its feasibility study in 2018. By that time it is also important that the MICE experiment is completed. The CDR/TDR process could then take place lasting probably at least 5 years to which the approval process needs to be added. The cost is unknown. The committee shall not venture a guesstimate of a date for muon colliders in this report. An update of the muon collider timeline by the MAP collaboration is expected to be available during Snowmass 2013.

### 5. Photon colliders

A photon collider can either be seen as an add-on to a linear collider, in which case one could contemplate operation sometimes in the active life (2030-2045) of the ILC. If a dedicated machine were to be built, a CERN time scale would probably place it starting construction in 2022 with possibly a somewhat faster timescale than TLEP because of the smaller tunnel, leading to the completion of 5 years of statistics sometimes between 2030 and 2035.